\documentclass[11 pt]{article}
\usepackage{lmodern}
\usepackage[margin=1in]{geometry}
\usepackage{booktabs}

\makeatletter
\renewcommand\section{\@startsection {section}{1}{\z@}%
                                   {-2.5ex \@plus 0ex \@minus 0ex}%
                                   {0.1ex \@plus0ex}%
                                   {\fontsize{15}{15}\bfseries}}
\renewcommand\subsection{\@startsection{subsection}{2}{\z@}%
                                     {-2.5ex\@plus 0ex \@minus 0ex}%
                                     {0.1ex \@plus 0ex}%
                                     {\fontsize{13}{13}\bfseries}}
\renewcommand\subsubsection{\@startsection{subsubsection}{3}{\z@}%
                                    {-2.5ex\@plus 0ex \@minus 0ex}%
                                     {0.1ex \@plus 0ex}%
                                     {\fontsize{11}{11}\bfseries}}
\makeatother
\usepackage{graphicx}
\usepackage{enumerate}
\usepackage{tikz}
\usetikzlibrary{decorations.pathreplacing,calligraphy}
\usepackage{url}
\usepackage{amsfonts}
\usepackage{latexsym}
\usepackage{tabularray}
\usepackage[tbtags]{amsmath}
\usepackage{amsmath,amssymb,graphicx}
\usepackage[ruled,lined]{algorithm2e}
\usepackage{color}
\usepackage{subcaption}
\usepackage{placeins}
\usepackage[skip=2pt]{caption}
\usepackage{graphics}
\usepackage{authblk}
\usepackage{xr}
\usepackage{float}
\usepackage[dvipsnames]{xcolor}
\usepackage[pdfborder={0 0 0}]{hyperref}
\usepackage{algpseudocode}
\usepackage{lipsum}
\usepackage{tablefootnote}
\usepackage[inline]{enumitem}
\usepackage{tabto}
\usepackage{multirow}
\usepackage{multicol}
\usepackage[flushleft]{threeparttable}
\usepackage[font=singlespacing]{caption}
\usepackage{tabulary}
\usepackage{todonotes}
\usepackage{tcolorbox}
\usepackage{wrapfig}
\usepackage{natbib}
\usepackage{setspace}

\hypersetup{
    colorlinks=true,
    linkcolor=black,
    filecolor=magenta,      
    urlcolor=blue,
    citecolor=blue,
}

\newtheorem{proof}{Proof}

\newtheorem{proposition}{Proposition}
\newcounter{remarkcnt}
\allowdisplaybreaks
\begin{document}

\setcounter{footnote}{1}

\title{ \vspace{-1cm} \LARGE
\textbf{A Multi-echelon Demand-driven Supply Chain Model for Proactive Optimal Control of Epidemics: Insights from a COVID-19 Study}
\\
}

\author[1]{Kimiya Jozani}
\author[1]{Nihal A. Sageer}
\author[2]{Hode Eldardiry}
\author[1]{Sait Tunc}
\author[1]{Esra Buyuktahtakin Toy}
\vspace{-0.5em}
\affil[1]{\footnotesize Grado Department of Industrial and Systems Engineering, Virginia Tech, Blacksburg, Virginia, USA \vspace{0em}}
\affil[2]{\footnotesize Department of Computer Science, Virginia Tech, Blacksburg, Virginia, USA }

\vspace{-3cm}

\date{}
\maketitle

\begin{abstract}
    {%
\textbf{Problem Definition:} 
Timely and effective decision-making is critical during epidemics to reduce preventable infections and deaths. This demands integrated models that jointly capture disease dynamics, vaccine distribution, regional disparities, and behavioral responses. However, most existing approaches decouple epidemic forecasting from logistics planning, hindering adaptive and regionally responsive interventions. \textbf{Methodology/Results:} 
We propose a novel epidemiological-optimization framework that jointly models epidemic progression and a multi-scale vaccine supply chain. The model incorporates spatio-temporally varying effective infection rates to reflect regional policy and behavioral dynamics. It supports coordinated, data-driven decision-making across spatial scales through two alternative formulations: (1) a multi-objective Gini-based model balancing efficiency and equity, and (2) a knapsack-based model that leverages regional vulnerability indicators for tractability and improved mitigation. To address computational complexity, we design two scalable heuristic decomposition algorithms inspired by the Benders decomposition. The model is validated using COVID-19 data from all 50 U.S. states and their counties. We introduce SARIMA-based forecasting as a novel approach for validating epidemic-optimization models under data limitations. The results show that our approach can prevent more than 2 million infections and 30,000 deaths in just six months while significantly improving the accessibility of vaccines in underserved regions. \textbf{Managerial Implications:} 
Our framework demonstrates that integrating epidemic dynamics with vaccine logistics leads to superior outcomes compared to traditional myopic policies. Epidemiologically optimized vaccine allocations diverge from real-world distributions, revealing that supply availability alone is insufficient. Equitable allocation not only enhances fairness, but also improves overall efficiency by prioritizing the most vulnerable populations, leading to better long-term public health outcomes. The model offers policymakers a scalable and operationally relevant tool to strengthen preparedness and ensure a more effective and equitable response to epidemics.

} 
\end{abstract}






\onehalfspacing
\maketitle

\vspace{-0.5cm}
\section{Introduction}


Epidemics have historically posed profound challenges to societies worldwide, significantly disrupting public health, economies, and social stability. In the 21st century alone, they account for approximately 14--20\% of annual deaths, with respiratory infections alone representing around 12\% \citep{owid-causes-of-death, owid-causes-of-death1}. Economically, epidemics such as the 2014 Ebola outbreak have resulted in losses that exceed \$53 billion \citep{huber2018economic}. Timely, data-driven decision making in epidemic response is essential but complicated by uncertainty, logistical constraints, and dynamic human behavior \citep{arifouglu2012consumption}. Consequently, there is an urgent need for integrated operational models that effectively address the complex dynamics of resource allocation during epidemics.\looseness-1

Among available public health interventions, vaccines remain among the most effective tools for controlling infectious diseases, capable of significantly curbing disease transmission and mortality when swiftly and strategically deployed \citep{yin2022risk}. However, synchronization of vaccine availability with the evolving dynamics of disease spread remains challenging \citep{braveman2006health, Sharomi201755}. This misalignment can severely hamper vaccination efforts, particularly when the spread of the disease outpaces the distribution of the vaccine \citep{kuzdeuov2020network}. Moreover, fluctuating demand requires that vaccine allocation strategies balance both reactive and proactive management approaches, further complicated by factors such as public perceptions and vaccine hesitancy \citep{moghadas2021impact}. Any mismatch between supply and demand can lead to shortages, surpluses, and even wastage, jeopardizing epidemic control \citep{paul2022mathematical,ozaltin2014optimal}. The COVID-19 pandemic underscored these difficulties, where the misalignment between vaccine supply and demand resulted in wastage rates as high as 20\% \citep{lazarus2022covid}.\looseness-1

Furthermore, social and ethical implications further complicate vaccine allocation, with disparities in vaccine access raising ethical concerns, deepening health disparities, and exacerbating the strains on healthcare systems \citep{braveman2003defining}. For example, socioeconomically disadvantaged populations were disproportionately affected due to inequitable resource distribution during COVID-19 \citep{mcgowan2022covid}; Uninsured and impoverished populations were 15 times more likely to die of infection due to lack of access to treatment resources \citep{fielding2020social}. Addressing these disparities is essential not only for protecting vulnerable groups, but also to mitigate the spread of the broader community by minimizing economic hibernation, thus reducing socioeconomic disruptions \citep{lane2017equity}. 
\looseness-1

Despite the recognized need for equitable and dynamic resource allocation, existing epidemic models often fail to capture essential real-world complexities. Traditional compartmental epidemic models typically treat vaccine supply chains separately or rely on simplified static assumptions, failing to integrate the multilayered logistical realities that span federal, state, and intra-state \citep{Gupta20221,Pathak20243999}. Moreover, although fairness and equity considerations have been conceptually recognized, their explicit integration into operational epidemic management frameworks remains limited and inadequately characterized \citep{bambra2022pandemic,de2025equity}. The literature notably lacks comprehensive mathematical frameworks that integrate dynamic epidemic conditions, logistics, and equity simultaneously, especially for large-scale implementations.

These critical gaps motivate our research, which aims to rigorously answer three pivotal questions: 
\begin{enumerate*}[label=(\roman*)]
    \item How can vaccine supply chain decisions across multiple spatial scales be effectively integrated with epidemiological dynamics and explicit equity considerations? 
    \item How can such an integrated framework be practically developed and scaled for implementation at a national level, such as in the United States?
    \item How do the strategies derived from the proposed integrated framework perform relative to the actual vaccine allocation policies?
\end{enumerate*}

We address these questions by developing a novel integrated optimization framework that dynamically aligns multi-scale vaccine logistics with epidemiological modeling and equity objectives. Our framework explicitly captures spatial and temporal heterogeneity in disease spread, resource availability, and public demand, improving responsiveness to real-world conditions. We further employ computationally efficient decomposition algorithms, enabling scalability and practical implementation. In doing so, we offer practical and methodological advancements, bridging significant gaps in the epidemic resource allocation literature, and providing robust decision-support tools to improve outcomes in future public health emergencies. In the following subsections, we provide an in-depth review of the literature highlighting existing shortcomings and clearly outline our contributions, establishing the innovative and impactful nature of this research.
\looseness-1

\subsection{Literature Review}

Compartmental models have long guided epidemic research by capturing the essential dynamics of disease transmission and intervention effectiveness \citep{brauer2019mathematical, kermack1927contribution, brauer2008mathematical, rabil2022effective, gandon2003imperfect}. Recent literature highlights the need to integrate logistical constraints with these epidemiological models to optimize resource allocation under real-world constraints such as limited vaccine supply, fluctuating demand, and evolving outbreaks \citep{yin2023covid, Comissiong20181841}. Although some studies have begun to address this integration \citep{Büyüktahtakın20181046, cocsgun2018stochastic}, many remain limited in scalability or human behavior aspects, leaving gaps for optimization-driven decision-making frameworks \citep{li2021mathematical}.\looseness-1

Equity, an essential yet underrepresented factor in the management of epidemics, significantly influences the effectiveness of resource allocation by mitigating disparities in access to healthcare care and outcomes~\citep{qiu2023impact}. As a multidimensional principle and ethical imperative, equity and fairness can be framed in various ways, from measuring disparities in health outcomes across populations to defining distribution strategies in terms of infection rates, capacity constraints, or resource access \citep{braveman2003defining, lane2017equity, sen2004health, braveman2006health, love2020methods, yin2021multi}. Although some research views equity and fairness as potentially detrimental to efficiency~\citep{bertsimas2011price}, recent studies demonstrate that efficiency and equity can coexist when properly integrated into resource allocation models \citep{delgado2022equity, xinying2023guide}.

Although some elements of the literature capture multiple dimensions of the pandemic response~\citep{ekici2014modeling,paul2022mathematical}, explicit considerations of dynamic treatment equity in the allocation of epidemic resources remain limited \citep{yin2021multi, delgado2022equity}, particularly within a comprehensive optimization framework that dynamically adapts to changing disease conditions and population needs \citep{Bennouna20231013, Pathak20243999}.


Addressing this critical gap, our research advances the state of the art by proposing a unified mathematical optimization framework that dynamically integrates disease progression, logistical constraints, and equity considerations, offering strategic insights into the effective and equitable management of epidemic response~\citep{sen2004health}. \looseness-1

\subsection{Contributions and Innovations}

Our methodological contributions are fourfold. First, to our knowledge, this is the first study to integrate epidemic dynamics and a multi-scale supply chain framework, encompassing state, intrastate, and pharmacy level allocations, into a unified optimization model, informing proactive policymaking during real-time disease progression. 
Unlike traditional methods that handle epidemiological modeling separately, our model optimizes public health outcomes directly through adaptive allocation decisions, addressing real-time epidemic progression and equity simultaneously. We present two innovative formulations: a Gini-based multi-objective model balancing efficiency and equity, and a knapsack-based model that prioritizes allocation using a composite vulnerability index.
\looseness-1

Second, 
we introduce spatially and temporally dynamic infection and vaccination rates within compartmental epidemic models. This innovation accurately captures the heterogeneity in transmission dynamics due to varying population behaviors, policies, and compliance levels, improving the realism and responsiveness of epidemic forecasts.
\looseness-1

Third, our knapsack-based formulation incorporates novel region-specific vulnerability indices derived from socioeconomic and epidemiological indicators, augmented by a max-min regularization approach to prevent inequitable allocations typical in conventional methods. We empirically demonstrate that our proposed method significantly outperforms traditional fairness metrics, both computationally and in epidemic control effectiveness.
\looseness-1

Finally,
addressing the computational challenges inherent in large-scale integrated models, we propose two tailored heuristic decomposition algorithms. Inspired by Benders decomposition \citep{benders1962partitioning, rahmaniani2017benders}, these algorithms exploit structural separability in temporal and spatial dimensions, significantly improving scalability and computational efficiency without compromising solution quality.
\looseness-1

From an application perspective, our contributions are twofold. First, we conduct the first comprehensive case study to optimize vaccine allocation across federal, state, county, and sub-county levels while accounting for complex dynamic epidemiological, logistical, and social vulnerability factors. We evaluate model solutions 
using real-world COVID-19 data across all 50 U.S. states and Washington, DC, optimizing vaccine allocation decisions at federal, state, county, and pharmacy levels.
Second, our adaptive decision-making framework utilizes real-time infection and demand data to strategically determine the optimal timing and placement of mass vaccination centers, dynamically aligning resource allocation with emerging epidemiological trends, thus reducing waste and enhancing operational responsiveness. Furthermore, we leverage the Seasonal Autoregressive Integrated Moving Average (SARIMA) model
for robust forecast validation, providing a novel integration of predictive analytics with optimization modeling in epidemic response management \citep{arunkumar2021forecasting, sah2023covid}.
\looseness-1

The remainder of the paper is organized as follows: \S\ref{Model} defines our integrated epidemic-logistic framework, explicitly outlining assumptions and model structure, with detailed mathematical formulations in \S\ref{model}. The methodological details are presented in \S\ref{method}. \S\S\ref{Case Study Data} and \ref{results} describe our numerical experiments, data calibration, and empirical findings. Finally, \S\ref{discussion} summarizes managerial insights and broader contributions.\looseness-1

\section{Integration of  Epidemic Dynamics and Vaccine Supply Chain}\label{Model}

\begin{wrapfigure}{r}{0.5\textwidth}
\centering
 \includegraphics[width=1\linewidth]{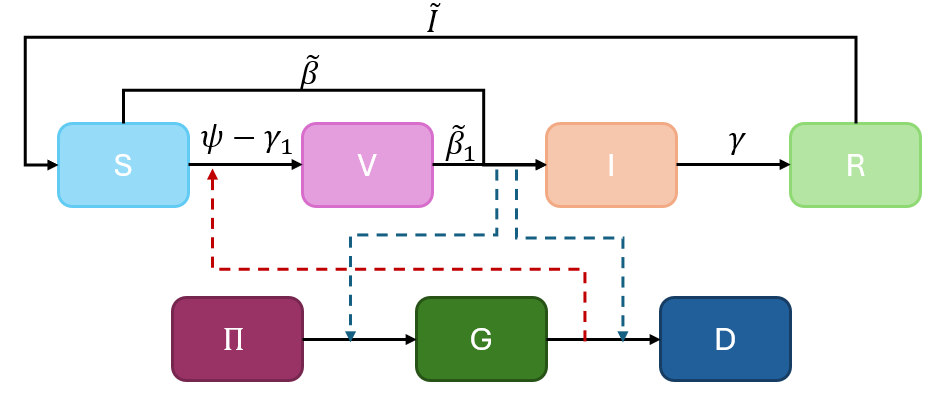}
 \caption{Proposed Policy-informed Integrated Epidemic-Resource Supply Chain Model}
 \label{fig:SVIR}
\end{wrapfigure}

In this section, we introduce our integrated epidemic-resource supply chain framework, highlighting a novel compartmental approach that explicitly integrates epidemic dynamics with resource logistics.
 We first describe the overall structure of the model and then detail its mathematical formulation, features, and key assumptions. 
Our proposed model is illustrated in Figure~\ref{fig:SVIR}, highlighting the integrated nature of epidemic progression and vaccine supply chain logistics. The upper part represents the classical compartmental structure (Susceptible--Vaccinated--Infected--Recovered (SVIR) model), while the lower section depicts the supply chain flow from vaccine manufacturers ($\Pi$) through distribution centers ($G$), and finally to the population demand ($D$).
Within the SVIR structure, susceptible individuals ($S$) transition to either vaccinated ($V$) or infected ($I$) compartments. Vaccination transitions depend explicitly on the availability of vaccines, determined by the supply ($G$), and the public's willingness or demand ($D$). This transition occurs at an immunization rate defined as $1/\gamma_1$. In contrast, susceptible individuals become infected based on factors such as social interactions and preventive measures. Recognizing the complexity of explicitly modeling these interventions, we introduce the concept of an \textit{effective infection rate} ($\tilde{\beta}$), which captures behavioral and policy factors indirectly and is discussed in detail in Section \ref{Case Study Data}.
To reflect scenarios where vaccine-induced immunity is incomplete or short-term, we let vaccinated individuals transition to the infected compartment at a reduced rate ($\tilde{\beta}_1$). The infected compartment subsequently transitions into the removed compartment ($R$), which aggregates individuals who recover or die from infection. Given our focus on reducing infection incidence, we do not model recovery and mortality separately. The  $R$  compartment aggregates both, with the assumption that only a fraction of \( R \) (i.e., the recovered) may re-enter the susceptible class. The transition rate from $I$ to $R$ is influenced by treatment resources and policies, with a removal rate corresponding to a recovery or death time ($\gamma$). The return transition from \( R \) to \( S \) reflects the possibility of reinfection at a rate $\sigma$ due to waning immunity among recovered individuals. 
In our integrated approach, vaccine allocation policies directly affect the transitions from $S$ to $V$, through the vaccination rate ($\psi - \gamma_1$), where $\gamma_1$ represents the immunization rate. The supply chain dynamics, from manufacturing ($M$) to distribution ($G$), depend on operational resources and policy directives. The flow from distribution ($G$) to demand ($D$) depends further on distribution policies and public vaccination willingness. In addition, infection and mortality data serve as critical input that influences allocation and administration decisions.\looseness-1

\subsection{Model Assumptions and Features}

To enhance practical applicability and clarity, the following assumptions are made in our model. These assumptions streamline our model implementation while ensuring realistic representation and operational feasibility, enabling focused analysis on vaccine allocation effectiveness and equity in epidemic management.
 \begin{enumerate*}[itemsep=-4pt,label=(Assumption-\arabic*),series=Assumptions ]
    \item Our framework exclusively considers vaccination as the primary intervention, reflecting middle-to-late-stage epidemic control strategies. Other interventions are implicitly captured via the effective infection rate ($\tilde{\beta}$). 
\item We assume a single-dose vaccination strategy to prioritize rapid coverage amid limited supply. Consequently, the second-dose compartment is excluded and the vaccine effectiveness parameters are calibrated accordingly. 
\item Reinfection among previously infected and vaccinated individuals is considered negligible within the short-term planning horizon (approximately six months), supported by empirical evidence of low reinfection rates \citep{nguyen2023sars}
\item Interregional migration and interactions are excluded to manage complexity. However, intra-regional (sub-regional) interactions remain explicitly modeled, aligning with existing regional data practices. This simplification is justified given current data limitations, but could be extended through more sophisticated interaction modeling frameworks \citep{Büyüktahtakın20181046, gutjahr2023fair}.
\item Vaccine efficacy and public vaccine preferences are uniform across regions and suppliers. This assumption reflects the early stages of vaccination, where obtaining any available vaccine takes precedence over specific brand preferences.
\item A multilayer supply chain model (manufacturer--region--subregion--local center) is assumed, explicitly depicted in Figure \ref{appendix-fig:supplychain} of Appendix~\ref{appendix: Appendix 1}. This structure suits decentralized resource control scenarios.
\item The primary objective of the model is to reduce infection rather than mortality, as infection rates are less influenced by external healthcare variables and reflect more directly the effectiveness of vaccination policies.
\end{enumerate*}
\looseness-1

\subsection{Mathematical Model Formulation}
\label{model}
We present two formulations to model an integrated epidemic supply chain with fair distribution of vaccines with limited resources: the Gini-based formulation (Section \ref{sec:form1}) and the Knapsack-based formulation (Sections \ref{sec:form2} and \ref{sec:decom2}). These formulations differ primarily in the incorporation of equity criteria. We first provide essential notation, followed by a clear and detailed description of each formulation. A comprehensive list of notations can be found in Appendix~\ref{appendix: Appendix 0}.
 
\subsubsection{Notation and Parameters}
The key indices in the formulations are as follows: $t$ denotes time periods, $j$ represents regions, $k_j, m_j, n_j$ refer to sub-regions of $j$, $l^k_j$ corresponds to local pharmacies in $k_j$, $o_j$ denotes potential mass vaccination sites in $j$, and $i$ indexes vaccine suppliers. \looseness-1

We have two different groups of parameters: epidemic and supply chain.
Epidemic parameters include: population's natural recruitment rate  ($\mu$), effective infection rate in unvaccinated ($\tilde{\beta}_{j}^{t}$) and vaccinated population ($\tilde{\beta}_{1,j}^{t}$), immunization rate ($\gamma_1$), removal rate ($\gamma$), reinfection rate ($\tilde{\sigma}^{t}$), immunity duration ($t_r$), indicator of the existence of long-term vaccine-induced immunity($ \psi \in \{0,1\} $), vaccination-eligible population in sub-region $k_j$ and region $j$ ($N_{k_j}$, $N_{j}$), and initial infected, susceptible to reinfection, and removed population ($I_j^0$ $\tilde{I}_j^0$,$R_j^0$) in region $j$. The supply chain parameters are as follows: 
\begin{enumerate*}[label=(\roman*)]\item
Monetary parameters: the unit cost of opening center, vaccine administration, vaccine per dose from $i$, transportation from $i$, transportation to $k_j$, transportation to $l_j^k$, inventory at $j$, inventory at $k_j$, inventory at $l_j^k$ ($c_{o_j}$, $c_j^t$, $g^{t}_{i}$, $g^{1,t}_{i,j}$, $g^{2,t}_{j,k_j}$, $g^{3,t}_{k_j,l_j^k}$, $w^{1,t}_{j}$, $w^{2,t}_{k_j}$, $w^{2,t}_{k_j}$, and $w^{3,t}_{l_j^k}$, respectively). We also have a national budget available ($B$).
\item Temporal parameters: the lead time to open a mass vaccination center, distributing vaccine from supplier $i$ to region $j$, distributing from region $j$ to sub-region $k_j$, and distributing from sub-region $k_j$ to local pharmacy $l^k_j$ ($l^0$, $l^1_{i,j}$,$l^2_{j,k_j}$, $l^3_{k_j,l^k_j}$, respectively).
\item Efficiency parameters at each time $t$: the vaccination capacity at sub-region $k_j$, the capacity of mass vaccination center $o_j$, the production capacity of supplier $i$, the vaccine wastage percentage, and the expected vaccine demand in region $j$. ($\chi^t_{k_j}$, $\kappa_{o_j}^t$, $\Pi_i^t$, $\xi$, $D_j^t$, respectively).
\end{enumerate*}

The model has four types of variables: \begin{enumerate*}[label=(\roman*)]
    \item compartmental variables ($S_{j}^{t}$, $I_{j}^{t}$, $V_{j}^{t}$, $R_{j}^{t}$, and $\tilde{I}_{j}^{t}$): previously discussed in Section \ref{Model}; \item linking variables: indicating the number of vaccines administered to the susceptible population in region $j$ and sub-region $k_j$ ($\Psi_{j}^{t}$ and $\Phi_{k_j,j}^{t}$) and to the recovered population ($\Xi_{j}^{t}$ and $\Omega_{k_j,j}^{t}$) in region $j$ and sub-region $k_j$, at each time $t$;
    \item center opening decision variables:  $\Upsilon_j^{I,t}$ and $\Upsilon_j^{D,t}$ as indicators of exceeding the infection and demand thresholds in the region $j$ according to the variable $\tau_j^t$ as the infection threshold, $X_{o_j}^t$  as a binary decision to open the center $o_j$, at time $t$;
    \item supply-chain variables ($G^{1,t}_{i,j}$, $G^{2,t}_{j,k_j}$, and $G^{3,t}_{k_j,l^k_j}$) and inventory variables ($W^{1,t}_{j}$, $W^{2,t}_{k_j}$, and $W^{3,t}_{l^k_j}$): showing different layers of supply chain as discussed in Appendix~\ref{appendix: Appendix 1}; and finally \item equity variables:  $u_{k_j}$ as vaccines per capita in sub-region $k_j$, $v_{m_j,n_j}$ as the absolute difference between $u_{m_j}$ and $u_{j,n}$, $\bar{u}_j$ as the average of $u_{k_j}$, $G(u_j)$ as the Gini coefficient in region $j$, $\eta$ as the maximum Gini in all regions, and $\zeta_t$ as the minimum per capita allocation across regions at time $t$.
\end{enumerate*} 
\looseness-1

\subsubsection{Gini-based Formulation}\label{sec:form1}
The Gini-based formulation, defined by Equations~\eqref{objective1-eq1.1}--\eqref{eq6.10}, simultaneously balances epidemic containment with both baseline vaccine equity and distributional fairness across regions. Specifically, the objectives are to minimize infections \eqref{objective1-eq1.1}, maximize the minimum per capita vaccination \eqref{objective2-eq1.2}, and minimize the maximum Gini coefficient representing the inequality in the vaccine distribution \eqref{objective3-eq1.3}. The constraints fall into five primary groups: epidemic simulation, dynamic decision-making thresholds, supply chain logistics, equity enforcement, and feasibility conditions, as presented below.\looseness-1

\vspace{-1cm}

\noindent{\footnotesize
\begin{multicols}{3}
\begin{subequations}

\begin{flalign}
\min   \sum_j \sum_t\big( \tilde{\beta}_{j}^t S^{t}_{j}I^{t}_{j}+\tilde{\beta}_{j}^{1,t} V^{t}_{j} I^{t}_{j} \big)\label{objective1-eq1.1}
\end{flalign}

\begin{flalign}
\max    \sum_{t} \zeta_{t}\label{objective2-eq1.2}
\end{flalign}

\begin{flalign}
\min   \eta \label{objective3-eq1.3}
\end{flalign}
\end{subequations}
\end{multicols}
\vspace{-0.5cm}
}

{\footnotesize
\noindent
\begin{subequations}
\begin{tikzpicture}[baseline={(current bounding box.center)}]
  \node (eqs) at (0,0) {
    \begin{minipage}{15.5cm}
\vspace{-2em}
\begin{flalign}
&   S^{t+1}_{j} = S^{t}_{j}+\mu -\mu  S^{t}_{j} -\tilde{\beta}_{j}^t S^{t}_{j}  I^{t}_{j}-\Psi_{j}^{t-\frac{1}{\gamma_1}}+\tilde{I}^t_j &\forall  j, t
\label{eq2.2}
\\
&   V^{t+1}_{j} = V^{t}_{j}-\mu  V^{t}_{j}+\Psi_{j}^{t-\frac{1}{\gamma_1}}-\tilde{\beta}_{j}^{1,t} V^{t}_{j} I^{t}_{j}-\psi\gamma_1 V^{t}_{j}& \forall  j, t
\label{eq2.3}
\\
&   I^{t+1}_{j} = I^{t}_{j}-\mu  I^{t}_{j}+\tilde{\beta}_{j}^t S^{t}_{j} I^{t}_{j} +\tilde{\beta}_{j}^{1,t} V^{t}_{j} I^{t}_{j}
-\gamma  I^{t}_{j} &\forall  j, t 
\label{eq2.4}
\\
&   R^{t+1}_{j} = R^{t}_{j}-\mu  R^{t}_{j}+\psi \gamma_1 V^{t}_{j}+\gamma I^{t}_{j} -\Xi_{j}^t &\forall  j, t 
\label{eq2.5}
\\
& \tilde{I}_j^t=\tilde{\sigma}^{t}  (\tilde{\beta}_{j}^{t} S^{t-t_r}_{j} +\tilde{\beta}_{j}^{1,t} V^{t-t_r}_{j}) I^{t-t_r}_{j}&\forall  j, t \geq t_r 
\label{eq2.6}\\
\label{eq6.1}
&S^0_{j}=N_{j}- I^0_{j}+\tilde{I}^0_j, V^t_{j}=0, R^t_{j}=R^0_{j}&t=0,\forall j
\end{flalign}

\vspace{-0.5cm}

    \end{minipage}
  };
  \draw[decorate, decoration={calligraphic brace, mirror, amplitude=10pt}, thick]
    ([yshift=4pt]eqs.north west) -- ([yshift=-4pt]eqs.south west);

  \node[rotate=90, anchor=center] at ([xshift=-0.7cm]eqs.west) {\textbf{Epidemic Simulation}};
\end{tikzpicture}
\end{subequations}

\noindent\begin{subequations}
\begin{tikzpicture}[baseline={(current bounding box.center)}]
  
  \node (eqs) at (0,0) {
    \begin{minipage}{15.5cm}
        \vspace{-0.5cm}
        \begin{flalign}
\label{eq20.0}
&\tau_j^t =
\begin{cases} 
\sum_{a=t-3}^{t-1} ((\tilde{\beta}_{j}^a S^{a}_{j} +\tilde{\beta}_{j}^{1,a} V^{a}_{j}) I^{a}_{j})/3 \qquad & \text{if } t \geq 3, \\
0.5\tilde{\beta}_{j}^{t}S_j^0 I_j^0 \qquad & \text{otherwise}
\end{cases}
&\forall  j, t
\\
&   (\tilde{\beta}_{j}^t S^{t}_{j} +\tilde{\beta}_{j}^{1,t} V^{t}_{j}) I^{t}_{j} \geq \tau_{j}^t - M  (1-\Upsilon_{j}^{I,t-l^0}) &\forall  j, t\geq l^0
\label{eq20.1}
\\
&   (\tilde{\beta}_{j}^t S^{t}_{j} +\tilde{\beta}_{j}^{1,t} V^{t}_{j}) I^{t}_{j} \leq \tau_{j}^t + M \Upsilon_{j}^{I,t-l^0} &\forall  j, t\geq l^0
\label{eq20.2}\\
& \sum_{k_j} \chi_{k_j}^t - D^{t}_{j}\leq  M  (1-\Upsilon_{j}^{D,t-l^0}) &\forall  j, t\geq l^0\label{eq20.3}
\end{flalign}

\vspace{-0.8cm}

\begin{flalign}
& \sum_{k_j} \chi_{k_j}^t - D^t_j \geq   -M \Upsilon_{j}^{D,t-l^0} &\forall  j, t\geq l^0\label{eq20.4}
\end{flalign}

\vspace{-0.8cm}

\begin{flalign}
&   X_{o_j}^t \leq \Upsilon_{j}^{I,t}+\Upsilon_{j}^{D,t} &\forall  j, t\geq l^0 \label{eq20.5}
\end{flalign}

\vspace{-0.8cm}

\begin{flalign}
& \sum_{o_j} X_{o_j}^t \leq 1 &\forall  j, t\geq l^0 \label{eq20.6}\\
& \sum_{o_j} X_{o_j}^t \geq \Upsilon_{j}^{I,t}-\frac{C}{B} &\forall  j, t\geq l^0 \label{eq20.7}
\end{flalign}

\vspace{-0.8cm}

\begin{flalign}
& \sum_{o_j} X_{o_j}^t \geq \Upsilon_{j}^{D,t}-\frac{C}{B} &\forall  j, t\geq l^0 \label{eq20.9}
\end{flalign}

\vspace{-0.8cm}

\begin{flalign}
&\Upsilon_j^{I,t}=\Upsilon_j^{D,t}=X_{o_j}^t=0&\forall t<l^0,j \label{eq20.8}
\end{flalign}

\vspace{-0.8cm}

\end{minipage}
  };
  \draw[decorate, decoration={calligraphic brace, mirror, amplitude=10pt}, thick]
    ([yshift=4pt]eqs.north west) -- ([yshift=-4pt]eqs.south west);

  \node[rotate=90, anchor=center] at ([xshift=-0.7cm]eqs.west) {\textbf{Dynamic Decision Making}};
\end{tikzpicture}
\end{subequations}

\noindent\begin{subequations}
\begin{tikzpicture}[baseline={(current bounding box.center)}]
  \node (eqs) at (0,0) {
    \begin{minipage}{15.5cm}
        \vspace{-0.5cm}
\begin{flalign}
& \sum_t (\Psi_{j}^t + \Xi_{j}^t) \leq N_j  &\forall j
\label{eq3.4}
\\
& \Phi_{k_j}^{t} + \Omega_{k_j}^{t} \leq \chi^t_{k_j} & \forall j,k_j,t
\label{eq3.5}
\\
& \Psi_j^{t}+\Xi_j^{t}\leq D_j^{t}  & \forall {t}, j 
\label{eq3.12} \\
& \Psi_{j}^{t} + \Xi_{j}^{t} = \sum_{k_j}(\Phi_{k_j}^{t} + \Omega_{k_j}^{t}) + (1-\xi)\sum_{o_j}\kappa^{t}_{o_j}  X_{o_j}^{t-l^0}& \forall  j, t\geq \min_{l_j^k,k_j}\{l^0,l^1_{i,j}+l^2_{j,k_j}+l^3_{k_j,l^k_j}\}
\label{eq3.6}
\end{flalign}

\vspace{-0.8cm}

\begin{flalign}
& \sum_{j}G^{1,t}_{i,j} \leq \Pi_i^{t} &\forall i, t
\label{eq3.7}
\end{flalign}

\vspace{-0.8cm}

\begin{flalign}
& \sum_{i}G^{1,t}_{i,j} + W^{1,t-1+l^1_{i,j}}_{j}=\sum_{k_j}G^{2,t+l^1_{i,j}}_{j,k_j} + (0.5+\frac{(t-l^0)}{2|(t-l^0)|})\sum_{o_j}\kappa^{t+l^1_{i,j}}_{o_j}  X_{o_j}^{t-l^0} +W^{1,t+l^1_{i,j}}_{j}&\forall j,t
\label{eq3.8}
\end{flalign}

\vspace{-0.8cm}

\begin{flalign}
&G^{2,t}_{j,k_j}+W^{2,t-1+l^2_{j,k_j}}_{k_j}=\sum_{l_j^k}G^{3,t+l^2_{j,k_j}}_{k_j,l_j^k}  +W^{2,t+l^2_{j,k_j}}_{k_j} &\forall k_j,j,t\geq l^1_{i,j}\label{eq3.9}
\end{flalign}

\vspace{-0.8cm}



\begin{flalign}
&(1-\xi)\sum_{l_j^k}(G^{3,t-l^3_{k_j,l^k_j}}_{k_j,l_j^k}  + W^{3,t-1-l^3_{k_j,l^k_j}}_{l^k_j})= \Phi_{k_j}^{t}+ \Omega_{k_j}^{t}+\sum_{l_j^k}W^{3,t-l^3_{k_j,l^k_j}}_{l^k_j} &\hspace{-0.9em}\!\forall k_j , j, t\geq l^1_{i,j}+l^2_{j,k_j}+l^3_{k_j,l^k_j}\label{eq3.10}
\end{flalign}

\vspace{-0.8cm}

\begin{flalign}
&C = \sum_t \bigg[ \sum_j c_j^t (\Psi_j^t+\Xi_j^t) \bigg]+ \sum_{i}\sum_{j}g^{t}_{i} G^{1,t}_{i,j} +\sum_{j}\sum_{i}g^{1,t}_{i,j} G^{1,t}_{i,j}+ \sum_{k_j}\sum_{j}g^{2,t}_{j,k_j} G^{2,t}_{j,k_j}  +\sum_{l_j^k}\sum_{j_k}g^{3,t}_{k_j,l_j^k} G^{3,t}_{k_j,l_j^k}
&\nonumber 
\end{flalign}

\vspace{-0.8cm}

\begin{flalign}&   +\sum_{j} \sum_{o_j} \max\{X_{o_j}^{t+1}-X_{o_j}^{t},0\} c_{o_j}^{t}  +\sum_{j}w^{1,t}_{j} W^{1,t}_{j}+ \sum_{k_j}w^{2,t}_{k_j} W^{2,t}_{k_j}  + \sum_{l_j^k} w^{3,t}_{l_j^k} W^{3,t}_{l_j^k} \leq B \label{eq3.11}&
\end{flalign}

\vspace{-0.8cm}



        
\begin{flalign}
\label{eq6.5}
&\Psi_j^t=\Xi_j^t= W_{j}^{1,t}=0  &\forall t < \min_{l_j^k,k_j}\{l^0,l^1_{i,j}+l^2_{j,k_j}+l^3_{k_j,l^k_j}\},  j
\end{flalign}

\vspace{-0.8cm}

\begin{flalign}
\label{eq6.6}
&\Phi_{k_j}^t=\Omega_{k_j}^t=G_{j,k_j}^{2,t}=W_{k_j}^{2,t}=G_{k_j,l_j^k}^{3,t}= W_{l_j^k}^{3,t}= 0 & \forall j,k_j,l_j^k, t<l^1_{i,j}+l^2_{j,k_j}+l^3_{k_j,l^k_j} 
\end{flalign}

\vspace{-0.8cm}

\end{minipage}
  };
  \draw[decorate, decoration={calligraphic brace, mirror, amplitude=10pt}, thick]
    ([yshift=4pt]eqs.north west) -- ([yshift=-4pt]eqs.south west);

  \node[rotate=90, anchor=center] at ([xshift=-0.7cm]eqs.west) {\textbf{Vaccine Supply Chain Model}};
\end{tikzpicture}
\end{subequations}

\begin{subequations}
\noindent\begin{tikzpicture}[baseline={(current bounding box.center)}]
  \node (eqs) at (0,0) {
    \begin{minipage}{15.5cm}
    
        \vspace{-0.6cm}
        
\begin{flalign}
&  u_{k_j}= \frac {\sum_t(\Phi_{k_j}^t+\Omega_{k_j}^t)}{N_{k_j}} &\forall k_j, j \label{eq4.1}
\end{flalign}

\vspace{-0.8cm}

\begin{flalign}
&  v_{m_j,n_j} \geq |u_{m_j}-u_{j,n}| &\forall m_j ,n_j 
\label{eq4.2}
\\ 
& \bar{u}_j = \frac{\sum_{k_j} u_{k_j}}{|K_j|} &\forall j
\label{eq4.4}
\\ 
& \sum_{m_j} \sum_{n_j} v_{m_j,n_j} = 2 G(u_j) \bar{u}_j (|{K_j}|)^2 & \forall j \label{eq4.5} \\
&   G(u_j) \leq \eta & \forall j \label{eq4.6}\\
& \frac{\Psi_j^{t}+\Xi_j^{t}}{N_j} \geq \zeta_{t}  & \forall {t}, j 
\label{eq4.0} 
\end{flalign}

\vspace{-0.8cm}
\end{minipage}
  };
  \draw[decorate, decoration={calligraphic brace, mirror, amplitude=10pt}, thick]
    ([yshift=4pt]eqs.north west) -- ([yshift=-4pt]eqs.south west);

  \node[rotate=90, anchor=center] at ([xshift=-0.7cm]eqs.west) {\textbf{Gini-Based Equity}};
\end{tikzpicture}
\end{subequations}

\begin{subequations}

\noindent\begin{tikzpicture}[baseline={(current bounding box.center)}]
  \node (eqs) at (0,0) {
    \begin{minipage}{15.5cm}
        \vspace{-1em}
        \begin{flalign}
&\Psi_j^t, \Xi_j^t, \Phi_{k_j}^t,\Omega_{k_j}^t, G_{i,j}^{1,t}, G_{j,k_j}^{2,t},G_{k_j,l_j^k}^{3,t},W_{j}^{1,t}, W_{k_j}^{2,t}, W_{l_j^k}^{3,t} \in \mathbb{Z^+}& \forall i, j,k_j,l_j^k,t\label{eq6.8}\\
\label{eq6.2}
&\Upsilon_j^{I,t}, \Upsilon_j^{D,t}, X_{o_j}^t \in \mathbb{B} &\forall t, j\\
\label{eq6.9}
&S_j^t, I_j^t,V_j^t, R_j^t,\tilde{I}_j^t,\tau_j^t, C,\zeta_t, u_{k_j}, v_{m_j,n_j}, G(u_j), \eta \in \mathbb{R^+}& \forall j,k_j,t\\
\label{eq6.10}
& t\in \mathbb{T},  j\in \mathbb{J}, i\in \mathbb{M}, m_j,n_j,k_j\in \mathbb{K}(j), o_j\in \mathbb{O}(j), l^k_j\in \mathbb{L}(k_j)
\end{flalign}

\vspace{-0.8cm}
\end{minipage}
  };
  \draw[decorate, decoration={calligraphic brace, mirror, amplitude=10pt}, thick]
    ([yshift=4pt]eqs.north west) -- ([yshift=-4pt]eqs.south west);

  \node[rotate=90, anchor=center] at ([xshift=-0.7cm]eqs.west) {\textbf{Sign \& Type}};
\end{tikzpicture}
\end{subequations}}

 \textbf{Epidemic Simulation Constraints.}
Equations \eqref{eq2.2}--\eqref{eq2.5} present the discretized epidemic simulation constraints \citep{liu2008svir, Büyüktahtakın20181046}.
Equation \eqref{eq2.2} recursively calculates the susceptible population for each region $j$.
 Instead of a fixed vaccination rate, we use $\Psi_{j}^t$ to represent the number of vaccine doses administered in region $j$ at time $t$, which better aligns with the supply chain modeling. This substitution allows for dynamic allocation in response to fluctuations in production, distribution, and demand.
 Constraint \eqref{eq2.3} is the simulation of the number of people vaccinated. The vaccinated population includes the term $\psi_{\gamma_1}V$, which indicates whether vaccine-induced immunity is partial or full. 
The number of infected population cumulates according to equation \eqref{eq2.4}. 
The removed population who recover or pass away from the diseases are shown in \eqref{eq2.5}. 
As mentioned in the assumptions, we do not consider the term $\Xi$ (recovered vaccinated) in the vaccinated compartment since natural immunity and vaccine-induced immunity significantly decrease the infection rate. The removed population comes from the full vaccine immunity, or either recovery or death from the infection.   
Equation \eqref{eq2.6} defines the population susceptible to reinfection at time $t$ as $\tilde{\sigma}^t$ times the infections that occurred $t_r$ periods earlier. Constraint \eqref{eq6.1} sets the initial epidemic conditions, with $I_j^0$ as the initial infections, $\tilde{I}_j^0$ as the initial re-infections in region $j$. Since vaccination is assumed to begin at time 0, all vaccination variables are set to zero at $t = 0$ for each region. Also, $R_j^0$ is the initial recovered or deceased population in region $j$. \looseness-1

\textbf{Dynamic Decision Making Constraints.}
Constraints \eqref{eq20.0}--\eqref{eq20.9} constitute infection- and demand-informed decisions to open mass vaccination centers.
Constraint \eqref{eq20.0} defines the infection threshold, set by policymakers. For $t \geq 3$, the threshold is the average number of infections over the past three periods; otherwise, it is 50\% of the initial period's infections. Constraints \eqref{eq20.1} to \eqref{eq20.4} define the epidemic and demand threshold conditions. Specifically, Constraints \eqref{eq20.1} and \eqref{eq20.2} use a binary indicator variable to determine whether the current number of infections in region $j$ during period $t$ exceeds a predefined threshold, $\tau_j^t$. Constraints \eqref{eq20.3} and \eqref{eq20.4} ensure that the existing capacity in region $j$ during period $t$ does not fall below the estimated regional demand for that period. Building on these constraints, as well as the budget constraint, Constraints \eqref{eq20.5} and \eqref{eq20.6} govern the dynamic decision-making process for opening mass vaccination centers. Constraint \eqref{eq20.5} requires that a center be opened if demand or infection thresholds are exceeded. Constraint \eqref{eq20.6} ensures that at most one mass vaccination center is opened per region, although this condition can be adjusted based on the specific requirements of the problem. In addition, constraints \eqref{eq20.7} and \eqref{eq20.9} are added to ensure center opening when needed and the budget is available, after a minimum period $l^0$ for center opening. The ratio $\frac{C}{B}$ approximates the feasibility of the budget, with values near 1 indicating a near-full utilization of the total available budget $B$, where $C$ includes all relevant costs as defined in Equation~\eqref{eq3.11}.
 \looseness-1

\textbf{Supply Chain and Epidemic-Logistics Linking Constraints.}
{Constraints \eqref{eq3.4}--\eqref{eq6.6} explicitly link epidemic dynamics to vaccine supply chain logistics. 
Constraints \eqref{eq3.4}--\eqref{eq3.9} correspond to the supply chain model illustrated in Figure~\ref{appendix-fig:supplychain} in Appendix~\ref{appendix: Appendix 1}. Constraints \eqref{eq3.4} ensure that the number of vaccines administered in a region does not exceed its population, consistent with a closed system assumption in which the population changes only through births and deaths (captured by the parameter $\mu$) and with the non-negativity of the stock variables $S$ and $R$. Constraints \eqref{eq3.5} and \eqref{eq3.12} account for the regional capacity and expected demand for vaccines at each time step. Specifically, \eqref{eq3.12} limits vaccine administration based on estimated demand, reflecting vaccine hesitancy and preventing excess supply; this constraint can be relaxed as the vaccination campaign progresses. Constraint \eqref{eq3.6} links epidemic-driven vaccine needs to supply-side variables by ensuring that the regional vaccine demand equals the sum of doses administered in its sub-regions and those administered via mass vaccination centers opened $l$ periods prior. Constraint \eqref{eq3.7} enforces supply limits by bounding each supplier’s outbound vaccines plus inventory. Constraint \eqref{eq3.8} models flow balance from suppliers to regions, while \eqref{eq3.9} ensures consistency between regional supply and sub-regional allocations. Constraint \eqref{eq3.10} ensures that planned vaccine administration at local pharmacies matches their available supply. Constraint \eqref{eq3.11} enforces the overall budget limit $B$, which includes costs for procurement, inventory, transportation, mass vaccination center setup, and administration. Unit transportation costs ($g^{1,t}_{i,j}$, $g^{2,t}_{j,k_j}$, $g^{3,t}_{k_j,l_j^k}$) may vary by distance or transportation mode. Initial supply chain conditions are specified in constraints \eqref{eq6.5} and \eqref{eq6.8}, ensuring that no vaccines are administered at the start of the planning horizon. Constraint \eqref{eq6.5} additionally ensures that vaccination at the state level can only begin once either a mass vaccination center is opened or doses are delivered to local pharmacies. Constraints \eqref{eq6.5} and \eqref{eq6.6} set initial supply and inventory levels to zero, accounting for lead times before vaccines become available. \looseness-1

\textbf{Equity Constraints.}  
{Constraints \eqref{eq4.1}--\eqref{eq4.0} quantify and enforce spatial and temporal equity. The Gini coefficient, linearly approximated through these constraints, directly targets equitable per capita vaccination coverage across all regions.}
 Constraints \eqref{eq4.1}--\eqref{eq4.6} enforce spatial equity through the Gini-based formulation \eqref{objective1-eq1.1}--\eqref{eq6.10}. The Gini coefficient is calculated as
\(
G(u) = \frac{1}{2 \bar{u} n^2} \sum_{m,n} |u_{m_j} - u_{j,n}|,
\)
where \( u_{m_j} \) and \( u_{j,n} \) represent per capita vaccine allocations in subregions \( m_j \) and \( n_j \) of region \( j \), \( \bar{u} \) is the average utility across subregions, and \( n \) is the total population of the region. We adopt the linearization method proposed by \cite{xinying2023guide} to integrate this measure into our model. In this formulation, constraint \eqref{eq4.1} defines the utility function, while constraint \eqref{eq4.6} minimizes the upper bound of the Gini coefficient nationwide. Finally, the variable domain, non-negativity, and integrality constraints are specified in \eqref{eq6.8}--\eqref{eq6.10}. Constraint \eqref{eq4.0} defines a lower bound on per capita vaccine allocation, introduced as a decision variable to be maximized across regions and time to promote temporal equity.\looseness-1

{\textbf{Feasibility and Non-negativity Constraints.} Constraints \eqref{eq6.8}--\eqref{eq6.10} establish the domain of decision variables, ensuring integrality and non-negativity of variables.}

\subsubsection{Knapsack-based Formulation}
\label{sec:form2}  
The knapsack-based formulation, given in \eqref{eq2.2}--\eqref{eq7.1}, reframes vaccine allocation as a resource prioritization problem, emphasizing equitable and efficient allocation to high-risk sub-regions. Unlike the Gini-based approach, this formulation resembles packing a relief truck: with limited vaccine supply, the goal is to serve the most at-risk sub-regions based on urgency and vulnerability. Here, sub-regional priority weights \( \delta_{k_j} \) are computed through a three-step process: (1) variable normalization, (2) priority aggregation, and (3) weight normalization. 
First, we normalize the Social Vulnerability Index (SVI)~\citep{svicounty}, initial infection rate \( \tilde{\beta}_0 \), and population for each sub-region \( k_j \) into Z-scores, denoted by \( d^{svi}_{k_j} \), \( d_{k_j}^{\beta} \), and \( d_{k_j}^p \), respectively. These components are aggregated into a composite priority score ($d_{k_j}$), which is then normalized to yield a relative allocation weight ($\rho_{k_j}$): {\footnotesize\(
d_{k_j} = \sqrt{(d^{svi}_{k_j})^2 + (d_{k_j}^{\beta})^2 + (d_{k_j}^p)^2},
\rho_{k_j} = \frac{d_{k_j}}{\sum_{k_j} d_{k_j}}.
\)} Normalization ensures that weights are summed to one within each region. The final weight is defined as \( \delta_{k_j} = (1 - A_j)\rho_{k_j} \), where \( A_j \) reflects healthcare access in region \( j \). The knapsack-based formulation is then presented as follows.\looseness-1

{\footnotesize
\begin{subequations}
\vspace{-0.6cm}
\begin{flalign}
&\eqref{objective1-eq1.1}, \quad \eqref{objective2-eq1.2}, \qquad 
\max\quad \sum_j \sum_t \bigg[ \sum_{k_j} \delta_{k_j}(\Phi_{k_j}^{t} + \Omega_{k_j}^{t}) + (1 - A_j) \sum_{o_j} \kappa^{t}_{o_j} X_{o_j}^{t} + \lambda \nu^j_t \bigg] &\label{objective3-eq2.1.3}
\end{flalign}

\vspace{-0.8cm}

\begin{flalign}
&\textrm{\textbf{s.t.}} \quad \text{\eqref{eq2.2} to \eqref{eq3.11} and \eqref{eq6.1} to \eqref{eq6.10}}, & \quad 
\frac{\Phi_{k_j}^{t} + \Omega_{k_j}^{t}}{N_{k_j}} \geq \nu^j_t \quad & \nu^j_t \in \mathbb{R}^+ \quad \forall t, k_j, j,\label{eq7.1}
\end{flalign}
\end{subequations}
}

\vspace{-0.2cm} 

The regularization term \( \lambda \nu^j_t \) in \eqref{objective3-eq2.1.3} ensures a baseline level of vaccine allocation even for low-priority sub-regions, penalizing solutions that exclude them entirely.  Equity constraints \eqref{eq7.1} include a lower bound on vaccines administered in each sub-region. This approach preserves the first two objectives of the Gini-based formulation (\eqref{objective1-eq1.1} and \eqref{objective2-eq1.2}), while replacing the Gini-based equity metrics with an informed priority-driven allocation rule, suitable for heterogeneous populations.\looseness-1

\section{Methodology}\label{method}

{Our model addresses a complex optimization problem involving nonconvex, mixed-integer, multiobjective quadratic formulations, reflecting the realities of large-scale vaccine distribution. Key challenges include nonlinear epidemic dynamics, integer decisions for facility openings, and conflicting objectives such as minimizing infections and ensuring equitable allocation. To manage nonlinear disease dynamics governed by SVIR equations \cite{liu2008svir}, we adopt a discretization-based simulation method \cite{Büyüktahtakın20181046}, converting continuous dynamics into tractable, discrete-time constraints. To reduce the computational burden of integrality constraints, we relax vaccine flow and administration variables to continuous nonnegatives, generating valid lower bounds \citep{wolsey2020integer}. For the multi-objective structure, we use scalarization and weighted sum approaches \cite{gunantara2018review, marler2004survey}, as formalized in equations~\eqref{eq:scale1} and~\eqref{eq:scale2}.}\looseness-1

\vspace{-0.5cm}

{\footnotesize
\begin{subequations}
\begin{flalign} \label{eq:scale1}
 \hspace{-0.5cm}& \max   \sum_t\big( \lambda_0 \sum_j\tilde{\beta}_{j}^t S^{t}_{j} I^{t}_{j}+\tilde{\beta}_{j}^{1,t} V^{t}_{j} I^{t}_{j} \big)+\lambda_{1,1} \sum_{t} \zeta_{t}-\lambda_{1,2} \eta \\
\label{eq:scale2}
 \hspace{-0.5cm}& \max \sum_t\big[\lambda_0 \sum_j(\tilde{\beta}_{j}^t S^{t}_{j} I^{t}_j+\tilde{\beta}_{j}^{1,t} V^{t}_{j} I^{t}_{j})+\lambda_{2,1} \zeta_{t}
 +\lambda_{2,2}\sum_j \big(\sum_{k_j}\delta_{k_j}(\Phi_{k_j}^{t} + \Omega_{k_j}^{t})+(1-A_j)\sum_{o_j}\kappa^{t}_{o_j}  X_{o_j}^{t}+\lambda \nu^j_{t}\big)\big],
\end{flalign}
\end{subequations}
}

\vspace{-0.2cm}

Each term combines multiple objective functions with scalar weights ($\lambda_0, \lambda_{1,1}, \lambda_{1,2}, \lambda_{2,1}, \lambda_{2,2}, \lambda$), reflecting policymakers' trade-off preferences.  Due to differences in scale, we normalize the objectives \citep{gunantara2018review}, typically dividing vaccine distribution and infection-related terms by total regional populations to constrain values within the $[0,1]$ range. Although finding exact upper bounds can be NP-hard, reasonable approximations based on known maximum values prove effective. To assess robustness, we compare solutions obtained both with and without normalization.\looseness-1

Solving large-scale nonlinear mixed-integer problems remains computationally intensive, even with advanced solvers \citep{martin2012large}. Traditional methods, such as Benders decomposition perform well on linear problems but face difficulties with nonlinear cut generation \citep{rahmaniani2017benders, barzanji2020decomposition}. To address these challenges, we develop two tailored heuristic algorithms: (i) a spatio-temporal decomposition for the Gini-based formulation (Appendix~\ref{appendix: Appendix 3}), and (ii) a temporal decomposition for the Knapsack-based formulation. These heuristics decompose the problem into smaller, tractable subproblems using a multi-level framework, enabling scalable and high-quality solutions for national, state, and county-level vaccine distribution under uncertainty.\looseness-1

\subsection{Knapsack-based Decomposition} \label{sec:decom2}
We propose a Knapsack-based temporal decomposition algorithm to effectively solve large-scale, nonconvex MIP formulations inherent in vaccine allocation problems. The original Knapsack-based formulation \eqref{eq2.2}--\eqref{eq7.1}, is computationally challenging due to its nonlinearity, integrality constraints, and multi-objective structure. To manage these complexities, our algorithm employs a master–subproblem decomposition approach that iteratively solves temporal master problems coupled with corresponding subproblems. 
In each iteration, the temporal master problem (${P}_M^t$) allocates vaccines at the regional level for each time period $t$, focusing primarily on minimizing the prevalence of infection. The subproblems (${P}_S$) then allocate these regional-level decisions into subregional distributions, emphasizing equity considerations. Information about budget availability and inventory constraints obtained from sub-problems solutions is iteratively fed back into subsequent master problems, creating a dynamic and responsive allocation mechanism. Epidemic dynamics and logistical resource constraints introduce temporal coupling across periods, thereby influencing future decisions based on earlier allocations. This structure inherently prioritizes timely vaccine distribution, consistent with the dynamics modeled by compartmental frameworks such as the SVIR model.\looseness-1

{\footnotesize
\begin{subequations}
\vspace{-0.5cm}
\begin{flalign}
\textbf{Master Problem at Period $t$ ($ {P}_M^t$)} & \qquad\min \qquad \lambda_0\sum_j \big( \tilde{\beta}_{j}^t S^{t}_{j} I^{t}_{j}+\tilde{\beta}_{j}^{1,t} V^{t}_{j} I^{t}_{j}\big) -\lambda_{2,1} \zeta_{t}&\label{dec2-objective1-eq1.1}
\end{flalign}

\vspace{-0.5cm}

\textrm{\textbf{s.t.}}   
\eqref{eq2.2}, \eqref{eq2.3}, \eqref{eq2.4}, \eqref{eq2.5}, \eqref{eq2.6}, \eqref{eq20.0}, \eqref{eq20.1}, \eqref{eq20.2}, \eqref{eq20.3}, \eqref{eq20.4}, \eqref{eq20.5}, \eqref{eq20.6}, \eqref{eq20.9}, \eqref{eq3.12}, \eqref{eq4.0}

\vspace{-0.5cm}

\begin{flalign}
&  (1-\xi)^{-1}(\Psi_j^{t-1}+\Xi_j^{t-1}) \leq \sum_{k_j} \chi^t_{k_j} +\sum_{o_j}\kappa^{t}_{o_j}  X_{o_j}^{t} & \forall j, t\label{dec1-eq1}\\
& (1-\alpha_{t-1}^1) \frac{(\Upsilon_j^{I,t}+\Upsilon_j^{D,t})}{2} \leq X_{o_j}^{t} \leq \Upsilon_j^{I,t}+\Upsilon_j^{D,t} &\forall j,t \label{decom2-eq2}\\
& \sum_j (\Psi_j^{t}+\Xi_j^{t}) \leq (1-\alpha^1_t+\alpha^1_t \frac{{\min_j \sum_
t \frac{1}{T}\hat{c}_j^t}}{ B-\hat{C}}) (\sum_i\Pi_i^{t} + \sum_j W_j^{1,t-1})& \forall t \label{decom2-eq4}
\end{flalign}
\vspace{-0.5cm}
\end{subequations}}

The information on the available budget and inventory of the sub-problems (${P}_S$)~\eqref{dec2-objective3-eq2.1.3} is integrated back into the master problem through constraint~\eqref{dec1-eq1}. Specifically, constraint~\eqref{decom2-eq2} ensures that binary vaccine center-opening decisions are activated whenever demand exceeds capacity or when infection thresholds are surpassed, contingent upon the budget feasibility indicator~$\alpha^1_t$. If~$\alpha^1_t = 0$, the model enforces vaccine center activation when necessary; otherwise, the lower bound is not binding, and constraint~\eqref{decom2-eq4} restricts the decisions. Constraint~\eqref{decom2-eq4} defines the allowable range for vaccine allocation, distinguishing between budget-feasible ($\alpha^1_t = 0$) and budget-constrained ($\alpha^1_t = 1$) scenarios. In the former, vaccine allocation is limited by supply availability; in the latter, it is bounded by a risk-adjusted affordability estimate, given in \eqref{decom2-eq4}, 
where $\hat{C}$ represents the cumulative cost.  $\hat{c}_j^t$ denotes the estimated cost per unit for region~$j$ at time~$t$. $\hat{c}_j^t$ captures both infection intensity and logistic costs and is calculated as
{\footnotesize $\hat{c}_j^t = \left( \sum_i g_i^t \Pi_i^{t-1} \middle/ \sum_i \Pi_i^{t-1} \right) + \left( g^{2,t}_{j,k_j} + \sum_{k_j} g^{3,t}_{k_j,l_j^k} / |{K}_j| + w^{1,t}_j \right)$}.
The corresponding subproblem~\eqref{dec2-objective3-eq2.1.3} allocates vaccines at the subregional level based on decisions provided by the master problem, explicitly emphasizing equity:\looseness-1 

\vspace{-0.4cm}
{\footnotesize\begin{subequations}
\begin{flalign}\label{dec2-objective3-eq2.1.3}
&\textbf{Sub-problem (${P}_S$)} \max \quad \sum_j \big[\sum_{t} \sum_{j^k}\delta_{k_j}(\Phi_{k_j}^{t} + \Omega_{k_j}^{t})+(1-A_j)\sum_{t}\sum_{o_j}\kappa^{t}_{o_j}  X_{o_j}^{t}+\sum_{t}\lambda \nu^j_{t}\big] & 
\end{flalign}
\end{subequations}
}

\vspace{-0.2cm}

{\textbf{s.t.} \eqref{eq3.6}, \eqref{eq3.7}, \eqref{eq3.8}, \eqref{eq3.9}, \eqref{eq3.10}}

Constraint~\eqref{eq3.11}, which enforces budget feasibility in the original problem, is replaced by Constraint~\eqref{decom2-eq4} in the master problem. Although \eqref{decom2-eq4} appears in the master problem, it incorporates subproblem-derived quantities, such as the cumulative cost $\hat{C}$ and per-unit estimates $\hat{c}_j^t$. Thus, budget feasibility is dynamically informed by subproblem solutions. The binary variable~$\alpha^1_t$ in~\eqref{decom2-eq4} is updated based on feasibility conditions: {\footnotesize $B - \hat{C} \leq (1 - \alpha^1_t) M$ and $B - \hat{C} \geq -\alpha^1_t M$}, where $M$ is a sufficiently large constant. These conditions ensure a dynamic switch between supply-limited and budget-limited allocation. The algorithm ends when a predefined number of iterations is reached ($t \geq T+2$), or when $\hat{C}$ falls within the acceptable range $0.95B \leq \hat{C} \leq B$. The flowchart and implementation details are provided in Appendix~\ref{appendix: Appendix 3}.\looseness-1

Comparing the objective functions of the original problem and the Knapsack-based Decomposition formulation \eqref{dec2-objective1-eq1.1}--\eqref{dec2-objective3-eq2.1.3}, we observe that the decomposition separates the problem into two components: minimizing infections at the regional level and fairly maximizing vaccination in sub-regions, subject to epidemic dynamics and supply chain constraints. However, the decomposition is not a relaxation of the original problem, as it imposes additional bounds on the budget—bounds derived from the optimal solutions of the subproblems rather than being fixed a priori. In Proposition~\ref{appendix-prop1} (see Appendix~\ref{appendix: Appendix 3}), we formally examine the relationship between the optimal solution obtained from the Knapsack-based Decomposition and the feasibility of the original problem. \looseness-1

Building on the Knapsack-based Decomposition formulation \eqref{dec2-objective1-eq1.1}--\eqref{dec2-objective3-eq2.1.3}, the Gini-Based Decomposition \eqref{appendix-dec1-objective1-eq1.1}--\eqref{appendix-dec1-eq3} decomposes the master problem temporally. After providing an estimate of the cost of sub-regional allocation, sub-problems are decomposed spatially. The exact cost is calculated at this step and {optimality and feasibility} cuts will be added to the master problem if the budget is violated. A detailed explanation is provided in Appendix~\ref{appendix: Appendix 3}. \looseness-1

\section{Case Study, Data Calibration, and Effective Infection Rate} \label{Case Study Data}

Calibrating optimization models ensures their outputs align with real-world data, enhancing reliability and decision-making under uncertainty \citep{gupta1998toward, Yin2024covid, bushaj2023simulation}. This process involves data acquisition, integration, cleaning, aggregation, and adjustment. A detailed description including the calibrated values of the parameters is provided in Appendix~\ref{appendix: Appendix 2}.\looseness-1
\subsection{Estimating Infection Rates}\label{infrate}

\begin{wrapfigure}{r}{0.5\textwidth}
\vspace{-0.8cm}
\centering
\includegraphics[width=1\linewidth]{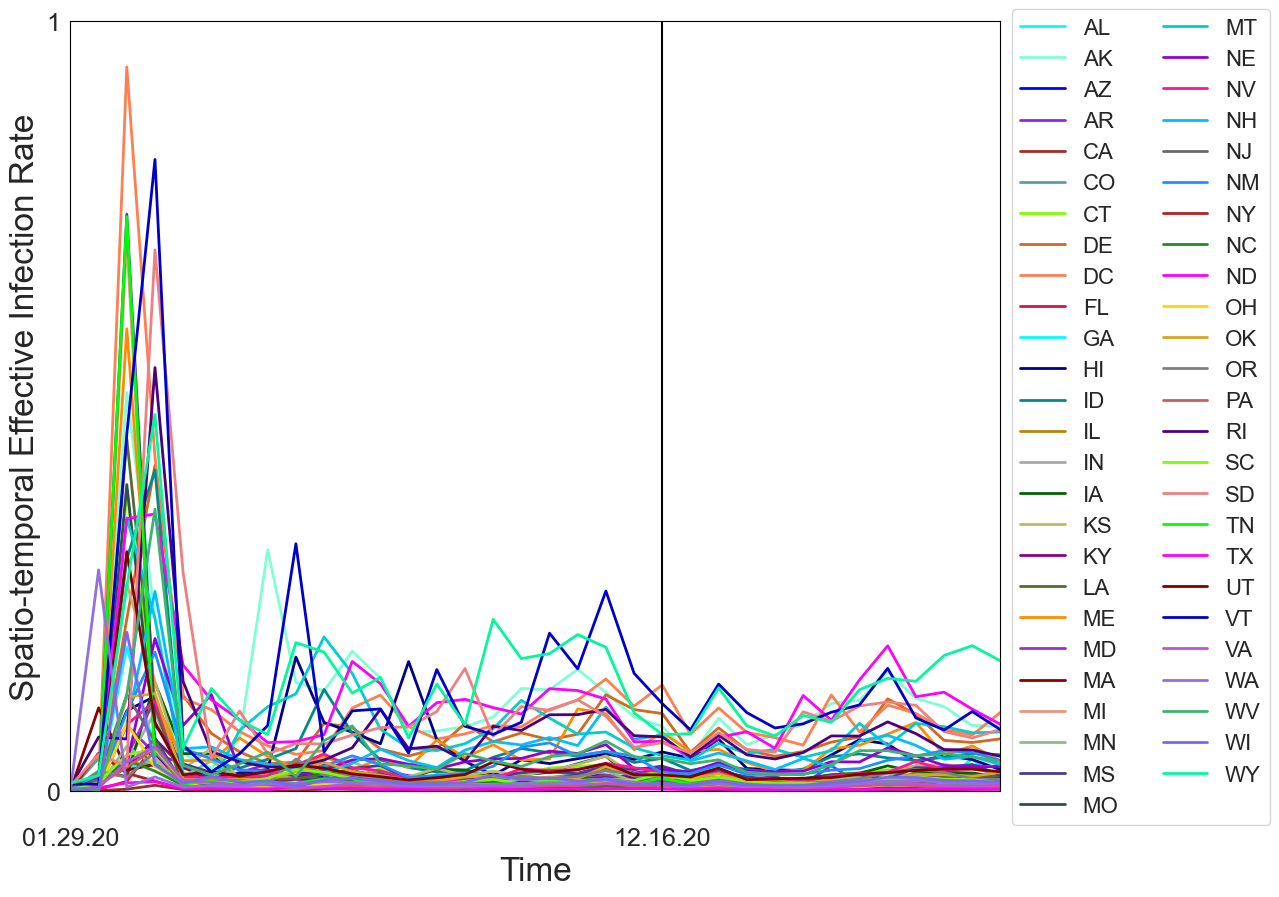}
\caption{\centering Temporal Effective infection rate in 50 states and DC}
\label{fig:infection rate}
\vspace{-0.8cm}
\end{wrapfigure}
This section describes the calibration of the effective infection rate $\tilde{\beta}_j^t$ for each region~$j$ over time. Although biological factors drive disease transmission, spatial and temporal variations in policy interventions (e.g., lockdowns, masking) significantly affect infection dynamics. To avoid overcomplicating the model with explicit behavioral compartments, we use an effective infection rate that implicitly captures contact-related variations. While our model currently assumes a single-dose vaccination framework, it can be extended to multiple doses by adding delay compartments, which would affect transitions to the infected state (see Figure~\ref{fig:SVIR}). Rather than using fixed vaccine efficacy, we empirically calibrate $\tilde{\beta}_j^t$ using case and vaccination data, reducing reliance on predefined vaccine effectiveness assumptions.
Figure~\ref{fig:infection rate} shows the spatio-temporal estimates of $\tilde{\beta}_j^t$ for the 50 U.S. states and the DC, derived using infection and vaccination data between January 21, 2020, and May 25, 2021. Our analysis period spans from start of public vaccination (December 13, 2020) to six months after vaccination rollout. These estimates align with available estimates of the effective reproductive number \citep{gostic2020practical, sy2021population, bugalia2023estimating, bokanyi2023real}. We calculate the effective reproductive number ($R_e$) using $R_e = \tilde{\beta}_j^t \gamma (1-P_j)$, where $P_j = S_j / N$ is the immune proportion of the population at time~$t$. The calibration procedure solves a linear system--an alternative representation of the SVIR dynamics (\eqref{eq2.2}--\eqref{eq2.4})--to get $\tilde{\beta}_j^t$ using the data. Specifically, we use the following system: \looseness-1

{\footnotesize
\begin{equation}\label{beta:matrix2}
\begin{bmatrix}
    S^{t}_{j}I^{t}_{j} & 1 & 0 \\
    \frac{\tilde{\beta}_{j}^{1,t}}{\tilde{\beta}_{j}^{t}}V^{t}_{j}I^{t}_{j} + S^{t}_{j}I^{t}_{j} & 0 & 0 \\
    \frac{\tilde{\beta}_{j}^{1,t}}{\tilde{\beta}_{j}^{t}}V^{t}_{j}I^{t}_{j} & 0 & 1
\end{bmatrix}
\cdot
\begin{bmatrix}
    \tilde{\beta}_j^t \\
    S_j^{t+1} \\
    V_j^{t+1}
\end{bmatrix}
=
\begin{bmatrix}
    (1-\mu)S_j^t + \mu - \Psi_j^{t-1} \\
    \text{cases}_j^t \\
    (1-\mu)V_j^t + \Psi_j^{t-1}
\end{bmatrix}
\end{equation}
}

\noindent Here, $\text{cases}_j^t$ represents the number of new infections in region~$j$ at time~$t$, inferred from the case data reported. The left-hand matrix includes known quantities from the previous time step ($S_j^t$, $I_j^t$, $V_j^t$), and the unknowns $\tilde{\beta}_j^t$, $S_j^{t+1}$, and $V_j^{t+1}$ are solved by matrix inversion. The initial conditions are set such that $I_j^0 = 1$ and $S_j^0 = N_j - 1$ to ensure epidemic seeding, while $V_j^0 = 0$. Vaccination data determine $\Psi_j^t$ (doses administered at time $t$) and $V_j^t$ (cumulative population vaccinated). The infection rate for the vaccinated population, $\tilde{\beta}_{j}^{1,t}$, is assumed to be 80\% lower than that of the unvaccinated, i.e., $\tilde{\beta}_{j}^{1,t} = 0.2 \cdot \tilde{\beta}_j^t$, consistent with estimates from~\citep{chen2022impact, moghadas2021impact}. Additionally, we account for 50\% underreporting in confirmed case data to adjust $\text{cases}_j^t$ during calibration. \looseness-1

\subsection{Computing Knapsack Weights}
\begin{wrapfigure}{r}{0.51\textwidth}
    \vspace{-1.5cm}
    \begin{center}
        \includegraphics[width=1\linewidth]{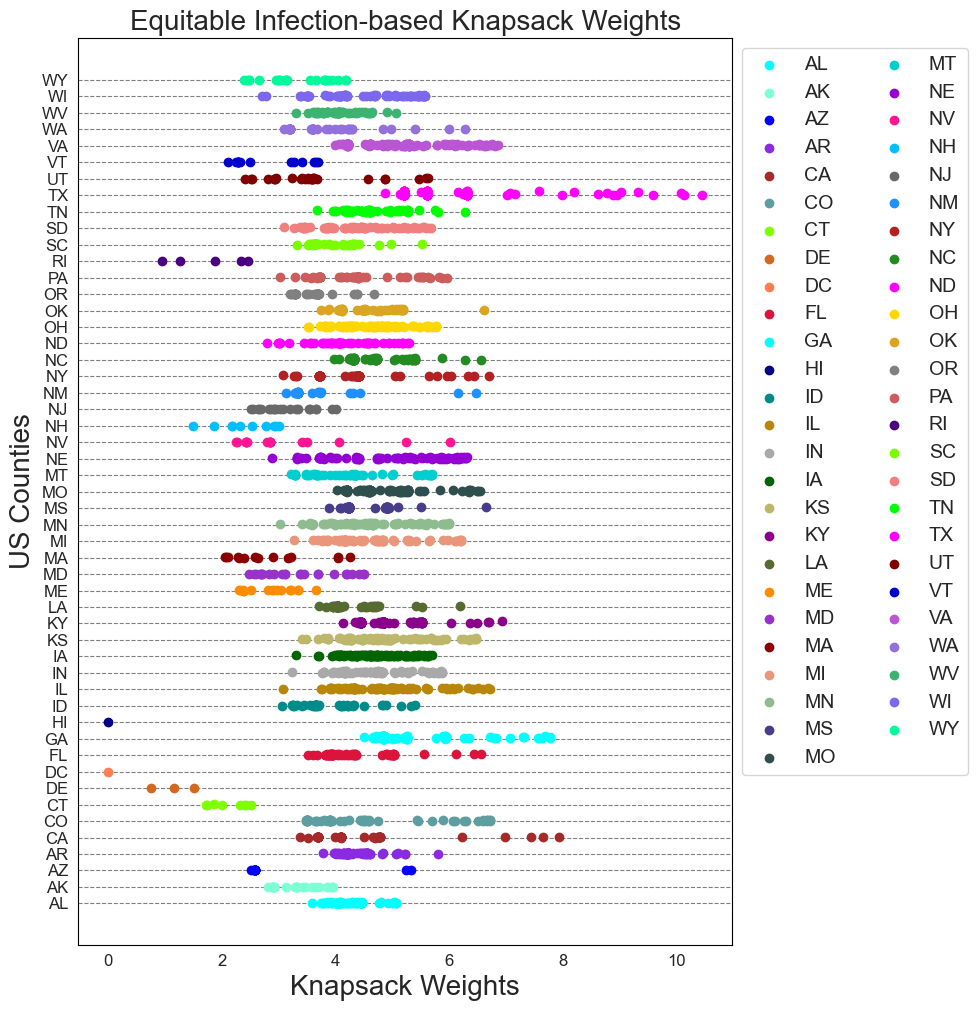}
    \end{center}
    \caption{\footnotesize County-level knapsack weights}
    \label{fig: county_knapsack}
    \vspace{-0.2cm}
\end{wrapfigure}

We define county-level weights \( \rho_{k_j} \) based on a composite score \( d_{k_j} \), as detailed in Section~\ref{sec:form2}. This composite metric captures and integrates the relative vulnerability based on SVI, infection rate, and population. For additional formulation details, see Section~\ref{sec:form2} and for more details, see Appendix~\ref{appendix: Appendix 2}. The resulting weights guide the allocation of vaccines by prioritizing counties with higher combined epidemiological and social risk. {Figure~\ref{fig: county_knapsack}} visualizes these logarithmic transformed weights. The higher values (right) correspond to the most vulnerable counties. States such as Texas, California, Florida, and New York exhibit broader distributions, underscoring the importance of county-level granularity in allocation decisions. \looseness-1

\subsection{Case Study, Model Validation and Generalizability}\label{sec: Validation}
Our case study includes 50 states of the United States and D.C. within the first six months of COVID-19 vaccination starting on 12/13/2020, as mentioned in Section \ref{Case Study Data}. Following model calibration, we apply two validation methods. First, assuming infection rates are available, we compare model-predicted infections with actual cases using observed infection rates, both nationally and by state over time. Second, when infection rates are unavailable, we estimate them using seasonal time series forecasting to validate future projections. 

\subsubsection{Model Calibration}\label{sec: generalvalidation}
The proposed optimization model relies on data to perform well in real-world settings. It is expected if calibrated properly, the model replicates the infected cases in reality when restricted to the actual resource allocation data \citep{Büyüktahtakın20181046}. For this purpose, we set the actual vaccine administered as the upper bound of the number of vaccines administered and compare the total number of infections with the actual data. An accurately calibrated model should select this bound because it reflects a known feasible solution. We set the upper and lower bounds of the vaccine administered as the actual values; in addition, we set the objective function coefficient as zero to make sure that it is not minimizing the infections but just checking the feasibility of our vaccine values and their resulting infections. The equity constraints are also relaxed. After calibrating all parameters, the total number of infections is 23,380,186, which is very close to the actual value of 23,380,197 (0.0001 percent less).
A two-tailed paired-$t$-test is used to analyze the difference between the pairs of predicted national new infections and the actual data in each period. All $p$ values for each period and each state after initiation are greater than 0.025, indicating that the projections of our model are statistically similar to the actual total number of infections for almost all periods.\looseness-1

\subsubsection{SARIMA for Validation} \label{sarima}
At the beginning of the vaccination period, when the public's compliance with social distancing policies is unknown, the data on the infection rate is unavailable. Validation under such circumstances is not as straightforward and requires some prediction. SARIMA models are one of the prediction tools used when the stationarity of the data is violated. This is the case during epidemics due to trends or seasonality characteristics of the observed cases and public compliance data. SARIMA uses past forecast errors and backshifts of the seasonal period in its predictions (\cite{hyndman20138}). Unlike machine learning models, which often require extensive feature engineering and large datasets, SARIMA explicitly incorporates seasonal and non-seasonal components through its parameters, offering a simple, transparent, and interpretable approach (\cite{hyndman2018forecasting}), well-suited for short- to medium-term forecasts. SARIMA's statistical foundation and diagnostic tools (e.g., ACF and PACF) allow for rigorous model evaluation and refinement, ensuring reliability in diverse applications (\cite{brockwell2002introduction}). Consequently, we use SARIMA in this paper to predict the infection rate using past data. A comprehensive discussion of our validation method is presented in Appendix~\ref{appendix: Appendix 4}. In brief, our results emphasize that model accuracy is both spatially and temporally heterogeneous, with higher reliability in the short term and in more stable regions.} This underscores the importance of frequent data acquisition for improving long-term forecasts. More advanced prediction techniques, such as neural networks, may be utilized to capture complex dynamics within the data. However, such methods require extensive data and computational resources to match or surpass the performance of well-calibrated statistical models.\looseness-1


\section{Results}\label{results}
 We implement our model on a desktop with the CPU model of Intel(R) Xeon(R) Gold 6326 CPU @ 2.90GHz, with 32 physical cores and 64 logical processors. Our solver is version 12.0.0 of Gurobi Optimizer, which has drastically improved in solving nonlinear MIP problems with more efficient CPU utilization compared to the previous versions. We use the MIP gap and Optimality gap metrics to assess the quality of the model's solution \citep{yilmaz2024expandable}. The MIP gap reports the solver's optimality bound at termination. The optimality gap is defined as $|\text{Heuristic Obj.} - \text{Incumbent Obj.}| / \text{Incumbent Obj.} \times 100$, where the Heuristic Obj. corresponds to the model's solution and the Incumbent Obj. is the best-known solution to the reference problem. Given the importance of the infection objective \eqref{objective1-eq1.1}, we use the term solution to refer to this objective.\looseness-1
\subsection{Results on Formulations and Decomposition Methodology}\label{subsec: results}

 We present numerical results in three parts: the Knapsack-based Formulation \eqref{eq2.2}--\eqref{eq7.1}, the Gini-based Formulation \eqref{objective1-eq1.1}--\eqref{eq6.10}, and a comparison of the two. We first provide the results of numerical experiments for the Knapsack-based formulation on the U.S. level. The complete Gini-based model proved intractable when solved at a national scale within a 12-hour solution time.  Hence, to assess the performance of our decomposition, we solve the knapsack-based problem for regions of New England (Maine, New Hampshire, Vermont, Massachusetts, Connecticut, and Rhode Island) and the Middle Atlantic (Delaware, Maryland, lower New Jersey, North Carolina, Pennsylvania, Virginia, and the District of Columbia), which in total have 12 states and a population of 57.6 million. The choice of these regions is illustrative rather than exhaustive. 
 To enable a comparative analysis, we use the same regions for all future analyzes, unless otherwise specified. \looseness-1

\subsubsection{Knapsack-based Formulation and Equity Results}\label{form2results}
This section details the numerical experiments for the Knapsack-based Formulation \eqref{eq2.2}--\eqref{eq7.1}. Table \ref{table: form 2-US} presents the computational performance and infection outcomes of the Knapsack-based formulation and the Knapsack-based decomposition for different scalarization methods and weight configurations across the United States, utilizing the scalarization method to select objective function weights (Section~\ref{method}).\looseness-1

The first column of Table \ref{table: form 2-US} is the methodology used. We simply use the term ``Scalarization'' to refer to the Knapsack-based Formulation \eqref{eq2.2}--\eqref{eq7.1} with scalarization in column 1. The second column is the objective weights that we use for scalarization in \eqref{eq:scale2}. The third column represents the number of infections caused by the selected method. The fourth column is the MIP gap provided by the solver. The fifth column is the optimality gap of the heuristic with respect to the solution in the first row. Finally, the sixth and seventh columns provide the solution time of each method and the infections averted by it, respectively, with respect to the actual infection data (23,380,197). \looseness-1

The weights in the second column represent a subset of various values tested empirically. Notably, the Knapsack Decomposition approaches achieved zero MIP gap $1000$~times faster, highlighting their efficiency compared to standard scalarization methods. Configurations using scalarization and normalization with weights proportional to the population and historical infection data in the 
$4th$~row of the table, resulting in the fewest infections with almost two million infections prevented, indicating strong performance in controlling the disease.  On the other hand, Knapsack Decomposition \eqref{dec2-objective1-eq1.1}--\eqref{dec2-objective3-eq2.1.3} solves the problem in seconds with 0 MIP gap, and over two million infections prevented. This produces a 0.07\% optimality gap. This shows that our decomposition scales well and shows superiority over the state-of-the-art solvers.
In summary, other scalarization methods with different weight settings showed varying levels of effectiveness and optimization gaps, reflecting the sensitivity of the results to the choice of weights and methodology.  \looseness-1


{\scriptsize
\begin{longtblr}[
    caption={\centering Results of Knapsack-based Formulation \eqref{eq2.2}--\eqref{eq7.1} for the Entire US, where objective weights respectively represent total infected cases, state lower bound, knapsack-based vaccine allocation, and county lower bound maximization}, 
    label={table: form 2-US}]{
  colspec={>{\centering\arraybackslash}p{4.2cm}
           >{\centering\arraybackslash}p{2.7cm}
           >{\centering\arraybackslash}p{1.5cm}
           >{\centering\arraybackslash}p{1.2cm}
           >{\centering\arraybackslash}p{1.2cm}
           >{\centering\arraybackslash}p{1.2cm}
           >{\centering\arraybackslash}p{1.2cm}}
}
\hline[1pt]
    Solution Methodology & Obj Weights $(\lambda_0,\lambda_{2,1},\lambda_{2,2},\lambda)$ & Infections (\# of cases)& MIP Gap (\%) & Opt Gap (\%) & Solution Time (s) & Infections Averted\\
 \hline
    Scalarized Formulation	&$(-1, 10^2, 10^{-3},10)$	&	21,515,219	&0.14	& 	-	&36,000	&	1,864,978	\\
    Scalarized Formulation	&$(-1, 10^2, 10^{-1},10)$	&	21,677,954	&0.83	& 	-	&24,000	&	1,702,243	\\
    Scalarized, Normalized Formulation  	&$(-\frac{1}{I^{0}+ R^{0} },\frac{20}{ N},\frac{1.4}{N},10)$	&	21,677,985	&0	& 	-	&863	&	1,702,212	\\
    Scalarized, Normalized Formulation 	&$(-\frac{2}{I^{0}+ R^{0} },\frac{20}{ N},\frac{1.4}{N},10)$	&	21,530,869	&0.21	& 	-	&5,500	&	1,849,328	\\
    Knapsack Decomposition 	&$(-1,1,1,10)$ 	&	21,691,092	& 0	& 	0.06	& 12.37	&	1,689,105	\\
    Knapsack Decomposition 	&$(-1,1000,1,10)$ 	&	21,083,982	& 0	& 	2.74	&12.92	&	2,296,215	\\
    Knapsack Decomposition 	&$(-1,\frac{N}{250},1,10)$ 	&	21,236,143	& 0	& 	2.1	&13.3	&	2,144,054	\\
    Scalarized, Normalized Decomposition	&$(-\frac{2}{I^{0}+ R^{0} },\frac{20}{ N},1,10)$	&	21,682,016	& 0	& 	0.70	& 12.68	&	1,698,181	\\
\hline[1pt]
\end{longtblr}
}








As a non-intuitive result, we observe that increasing the equity coefficient until its objective term overshadows the infection objective (meaning \eqref{objective1-eq1.1} is greater than minus \eqref{objective2-eq1.2}) effectively reduces infections. This suggests that placing more emphasis on equitable allocation can improve overall epidemic outcomes, at least under certain conditions. We hypothesize that, in such high-demand and low-supply settings, enforcing equity helps approximate the equal distribution across counties. Additionally, imposing equity discourages short-sighted and overly aggressive allocation to only the most severely affected areas, caused by the temporal structure of the decomposition. By promoting balance, equity can support more strategic long-term control of disease spread. \looseness-1
This balance between efficiency (here, controlling infections) and equity is particularly critical, as local maximization of short-term gains, achieved through a rolling-horizon approach and temporal decomposition, can lead to vaccine distribution disparities among states. These disparities, exacerbated by the dynamic nature of epidemics, can undermine the overall efficiency of the system. Incorporating equity into short-term decision-making can improve efficiency and address these disparities. Therefore, maintaining a strategic balance in periodic decisions and adopting long-term strategies are essential to improve the system performance, especially with similar demands and infection rates in subsystems. \looseness-1

Similar to Table \ref{table: form 2-US}, we have Table \ref{table: form 2-MANE region} presenting the computational performance and infection outcomes of the Knapsack-based formulation and the Knapsack-based decomposition for different scalarization methods and weight configurations in the Middle Atlantic and New England regions. Infections averted are calculated as the difference between observed cases ($4,588,856$) and those predicted by each model. The supply level is set to the actual allocations to these regions, which is less than the actual doses administered. A 20\% increase in the supply satisfies the actual demand for vaccines and decreases the infections to 3,911,523 cases. Similar to the US-level problem, the Knapsack-based decomposition is superior to the full formulation. The scale of efficiency is lower ($300$~times faster), which is expected due to the smaller scale of the problem. The optimality gap is $0.1\%$, which indicates the high performance of the decomposition method. \looseness-1

{\scriptsize
\begin{longtblr}[
    caption={\centering Results of Knapsack-based Formulation \eqref{eq2.2}--\eqref{eq7.1} for New England and Middle Atlantic Regions}, 
    label={table: form 2-MANE region}]{
  colspec={>{\centering\arraybackslash}p{4.2cm}
           >{\centering\arraybackslash}p{2.5cm}
           >{\centering\arraybackslash}p{1.5cm}
           >{\centering\arraybackslash}p{1.2cm}
           >{\centering\arraybackslash}p{1.2cm}
           >{\centering\arraybackslash}p{1.2cm}
           >{\centering\arraybackslash}p{1.2cm}}
}
\hline[1pt]
    Solution Methodology& Obj Weights $(-1,\lambda_{2,1},\lambda_{2,2},\lambda)$&Infections (\# of cases)& MIP Gap (\%) & Opt Gap (\%) & Solution Time (s) & Infections Averted\\
 \hline
    Scalarized, Normalized Formulation & $(\frac{2}{I^{0}+ R^{0} },\frac{-20}{ N},\frac{-1.4}{N},\frac{-1.4}{ N})$&4,162,840&0.05&-&660&426,015\\
    Knapsack Decomposition & $(\frac{2}{I^{0}+ R^{0} },\frac{-20}{ N},-1,-1)$&4,167,915&0&&0.1&420,940\\
\hline[1pt]
\end{longtblr}
}

In addition to the aforementioned experiments, we implement the problem for different regions within the US. Regional analysis reveals varying performance: in the Southwest, the problem is solved in 0.6 seconds with a 0.06\% MIP gap. However, in the Midwest, West, and South, the solver struggles, showing MIP gaps of 29.15\%, 10.98\%, and 25.8\% respectively within 9 hours; however, our decomposition still is able to produce high-quality results, in under a second. These findings highlight the effectiveness of the Knapsack-based Decomposition in solving the problem efficiently and minimizing infections close to the optimal solution.
\looseness-1

\subsubsection{Gini-based Formulation }

We also implement a set of numerical experiments for Gini-based Formulation \eqref{objective1-eq1.1}--\eqref{eq6.10} and the modified spatiotemporal heuristic decomposition (Gini-based Decomposition \eqref{appendix-dec1-objective1-eq1.1}--\eqref{appendix-dec1-eq3} for the Middle Atlantic and New England regions.
The Gini-based Formulation~\eqref{objective1-eq1.1}--\eqref{eq6.10} was solved in 3,600 seconds with a 0.04\% MIP gap. In contrast, the Gini-based Decomposition~\eqref{appendix-dec1-objective1-eq1.1}--\eqref{appendix-dec1-eq3} achieved a 0.06\% optimality gap in under 4 seconds. Both approaches yield a zero Gini coefficient, indicating full equity across states, as enforced by the objective of the decomposition sub-problem~\eqref{appendix-dec1-objective2-eq1.1}. The full formulation averted 392,254 infections, while the decomposition averted 419,270 infections--denoting a difference of 27,0165 cases. A more detailed result is provided in Appendix~\ref{appendix: Appendix 6}. Although solving the Gini-based Formulation \eqref{objective1-eq1.1}--\eqref{eq6.10} for the U.S. level is intractable, Gini-based Decomposition \eqref{appendix-dec1-objective1-eq1.1}--\eqref{appendix-dec1-eq3} can solve it with the total number of infections of 21,083,982 and a 0\% MIP gap in $239.2$ seconds. In addition, the maximum Gini coefficient \eqref{eq4.5} is 0.038, with a mode of 0.\looseness-1

\subsubsection{Comparative Analysis of Formulations}

Upon implementing the Gini-based and Knapsack-based models, we find that the Knapsack-based Formulation \eqref{eq2.2}--\eqref{eq7.1} demonstrates greater efficiency and effectiveness. Both the Gini-based and Knapsack-based Formulations yield similar results in terms of infections (4,162,840 and 4,196,602, respectively) and MIP gaps (0.05\% and 0.04\%, respectively). Furthermore, the Knapsack-based Formulation maintains a low Gini coefficient ranging from 0 to 0.0286, indicating its efficacy not only in reducing infections but also in achieving equity comparable to the Gini-based Formulation \eqref{objective1-eq1.1}--\eqref{eq6.10}. 
When comparing our model's outcomes with actual data, we observe varying behaviors when bounding vaccine administration by actual values. In the next section, we will compare the potential impacts of implementing the Knapsack-based decomposition with actual allocations. 
\looseness-1

\subsection{Evaluating Model-Based vs. Actual Vaccine Allocations}

\begin{figure}[t]
\vspace{-0.4cm}
    \centering
        \includegraphics[width=0.9\linewidth]{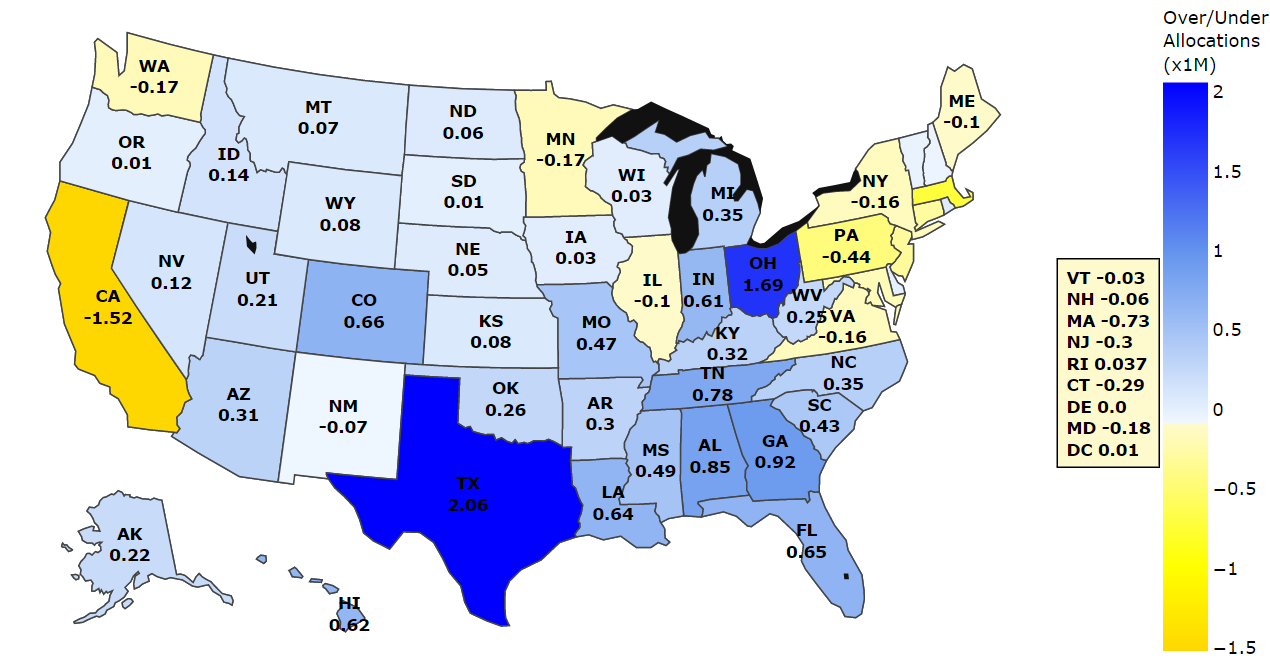}
        \caption{\centering\footnotesize Disparities between actual and model-based vaccine allocations under state-level caps, shown in millions of doses. Positive values (+) indicate overallocation (blue), and negative values (–) indicate underallocation (yellow).}
        \label{fig:map1}
\vspace{-0.5cm}
\end{figure}

We compare the first doses of vaccines allocated by the Knapsack Decomposition \eqref{dec2-objective1-eq1.1}--\eqref{dec2-objective3-eq2.1.3} with those allocated in practice ($\sum_i \sum_t G^{1,t}_{i,j}$). {Figure~\ref{fig:map1}} presents the model-based allocation per state with state-level caps, requiring vaccines to be limited by the actual numbers administered. This contrasts with actual distributions, which often deviate from need due to policy or logistical constraints. In this map, the shadings represent the difference between actual and model allocations, where blue and yellow mean over- and under-allocations. The numbers below the state abbreviations show the amount of over-allocation in thousand doses. The negative values show under-allocation.
As we can see, California is the most under-allocated state. On the other hand, Texas is over-allocated by $1$~million doses of vaccines, which could be used to cover the demand in California. It should be noted that there is a significant difference in the total number of vaccines administered and allocated in states such as California, Massachusetts, New York and Virginia, according to CDC data.
This observation indicates possible vaccine trips to get vaccinated, or some states have ordered more than allocated or used their second dose resources.
Massachusetts, Connecticut, and Vermont have the highest ratio of the difference between the model and actual to the actual allocations. States such as Alaska and Ohio may have received excess doses in reality due to lower demand. By enforcing state-level caps, we ensure that the reduction in infections is caused by improving allocations, rather than increased vaccination.\looseness-1

Solving the model without enforcing state-level upper bounds on the vaccines administered allows infection prioritization with relaxed bounds on the demand, allowing high-need areas to receive greater support. 
Comparing the figures for state-bounded and unrestricted administration reveals that incorporating state-level bounds improves alignment with actual distributions, but may constrain responsiveness to real-time needs. In both figures, we observe that the Middle Atlantic states with higher infection-demand scores (higher infection or willingness to receive vaccines) seem to be under-allocated in practice. On the other hand, states like California are also over-allocated; however, this is despite the high demand and due to mandatory stay-at-home and closures, which controlled the disease and made California's infection situation less urgent compared to other states. In contrast, states such as Wyoming have a relatively high infection score, which requires them to receive more vaccines when the upper bound is relaxed, indicating a low willingness to receive vaccines in these areas. CDC data confirm a mismatch between supply and demand, with some states receiving more vaccines than needed while others received less. Our model addresses this discrepancy, suggesting a more efficient and equitable allocation strategy that reduces cross-state travel for vaccination and helps limit disease transmission \cite{Gupta20221}. Furthermore, not imposing the limits caused a myopic allocation, which increased the number of infection cases by 1.3 million, indicating the need for long-term parity in the allocations. \looseness-1

\subsection{Sensitivity Analysis}\label{subsec: sensitivity}

We perform a local one-at-a-time (OAT) sensitivity analysis \citep{borgonovo2016sensitivity} to evaluate the sensitivity of the model to parameters, critical for decision-making under uncertainty \citep{tortorelli1994design}. We analyze six parameters: budget, supply, infection rate, vaccine effectiveness, infrastructure capacity, and demand. The Knapsack-based Decomposition model \eqref{dec2-objective1-eq1.1}--\eqref{dec2-objective3-eq2.1.3} with weights $(-1,\frac{N}{250},1,10)$ is used as a reference due to its computational efficiency.\looseness-1

{\scriptsize
\begin{longtblr}[
    caption={Sensitivity Analysis of Budget Levels}, 
    label={table:sensBudget}]{
  colspec={lp{2cm}p{1.2cm}p{2.5cm}p{3cm}p{2.5cm}}
}
\hline[1pt]
Budget Level &Budget (\$B)&Cost(\$B)& Infections (cases)&Vaccination (doses)&Centers Opened\\
\hline
Ample&	5		&	4.54	&	21,236,143	&	140,691,431	&	45	\\
Medium&	2.5		&	2.35	&	22,216,544	&	82,569,883	&	20	\\
Scarce&	1		&	0.82	&	32,485,816 	&	26,566,950	&	0	\\
\hline[1pt]
\end{longtblr}
}

We begin by examining economic constraints through three budget scenarios: sample, medium, and scarce, based on historical estimates. Table~\ref{table:sensBudget} summarizes the results, where each row corresponds to a budget scenario (ample, medium, or scarce). The columns report the total available budget (in billion USD), the model-estimated supply chain cost, cumulative infections, total vaccine doses administered, and the number of mass vaccination centers opened under each scenario.
 As shown in Table~\ref{table:sensBudget}, reduced budgets significantly increase infections by limiting vaccine availability and restricting the number of vaccination centers opened. With abundant resources, vaccine uptake is constrained by supply and demand, but cost feasibility becomes the binding constraint under a strict budget. The cost-budget gap, affected by average unit cost $\bar{c}_j^t$, may reflect a conservative estimation in \eqref{decom2-eq4}.\looseness-1

Table \ref{table:sensFull} presents the results of the sensitivity analysis for the parameters mentioned in the first column. The second column shows the change in the parameter value. The third and fourth columns, ``Infections'' and``Vaccination'', respectively, report the number of infected individuals and the number of vaccinated individuals in vaccination centers and pharmacies. The next column is ``Center Opened,'' which refers to the number of mass vaccine centers that the model decided to open in each scenario.
In the last two columns, we detail the model's sensitivity to changes in parameters, showing the number of infections averted compared to the actual 23.38 million cases and the variation in the number of vaccines administered relative to the actual 140.69 million first doses.\looseness-1

As we can see in Table~\ref{table:sensFull}, increasing supply reduces infections by one million, although it does not increase vaccinations, as demand remains the binding constraint. This reduction in infections under ample supply highlights the importance of early vaccination. To test the generalizability of the disease, we vary the infection rate. With only 20\% more infection rate, the cases increase by more than $100\%$, underscoring the sensitivity of the system to the nature of the disease. Vaccine effectiveness also affects outcomes. Although the total number of vaccines administered remains stable, the model opens more mass vaccination centers with lower effectiveness, consistent with capacity-related constraints (\eqref{eq20.1}--\eqref{eq20.6}). However, due to the widespread availability of local pharmacies and high demand, the model shows limited sensitivity to changes in center capacity, indicating that the current infrastructure is largely sufficient.\looseness-1

{\centering
{\scriptsize
\begin{longtblr}[
    caption={Sensitivity Analysis of Vaccine Supply, Infection Rate, Vaccine Efficacy, Capacity, and Demand}, 
    label={table:sensFull}]{
  colspec={lp{4.2cm}p{1.2cm}p{1.2cm}p{1.2cm}p{1.3cm}p{1.2cm}},
}
\hline[1pt]
Parameter & Amount &Infections (cases)&Vaccination (persons)&Centers Opened(No.)&Infections Change(\%)&Vaccination Change(\%)\\
\hline
Supply Level	&	20\% Less	(120.27 M Doses)	&	28,278,460	&	120,270,324	&	45	&	20.95	&	-14.51	\\
	&	Actual	(150.34 M Doses)	&	21,236,143	&	140,691,431	&	45	&	-9.17	&	0.00	\\
	&	20\% More	(180.41 M Doses)	&	20,348,583	&	140,691,431	&	30	&	-12.97	&	0.00	\\
Infection Rate	&	20\% Less	(Calculated in Section~\ref{infrate})	&	10,681,067	&	140,691,431	&	7	&	-54.32	&	0.00	\\
	&	Actual	(Calculated in Section~\ref{infrate})	&	21,236,143	&	140,691,431	&	45	&	-9.17	&	0.00	\\
	&	20\% More	(Calculated in Section~\ref{infrate})	&	45,797,591	&	140,691,431	&	107	&	95.88	&	0.00	\\
Vaccine Effectiveness	&	20\% Less	($\beta_1$=0.64$\beta$)	&	22,248,813	&	140691431	&	56	&	-4.84	&	0.00	\\
	&	Actual	($\beta_1$=0.8$\beta$)	&	21,236,143	&	140,691,431	&	45	&	-9.17	&	0.00	\\
	&	20\% More	($\beta_1$=0.96$\beta$)	&	20,497,047	&	140,691,431	&	45	&	-12.33	&	0.00	\\
Capacity	&	20\% Less	(Census Pharmacy Data)	&	21,236,143	&	140,691,431	&	46	&	-9.17	&	0.00	\\
	&	Actual	(Census Pharmacy Data)	&	21,236,143	&	140,691,431	&	45	&	-9.17	&	0.00	\\
	&	20\% More	(Census Pharmacy Data)	&	21,236,143	&	140,691,431	&	45	&	-9.17	&	0.00	\\
Demand	&	20\% Less	(112,553,145 Persons)	&	22,240,835 &	112,553,145	&	45	&	-4.87	&	-20.00	\\
	&	Actual	(140,691,431 Persons)	&	21,236,143	&	140,691,431	&	45	&	-9.17	&	0.00	\\
	&	20\% More	(168,829,717 Persons)	&	21,213,853	&	150,337,905	&	33	&	-9.27	&	6.86	\\
\hline[1pt]
\end{longtblr}
\noindent\tiny{\footnotemark[1],\footnotemark[2]{Data sources available in Appendix~\ref{appendix: Appendix 2}}
}}}

The demand for vaccines is a critical factor in supply chain management, especially evident in vaccination campaigns where individual choices significantly influence the results. As shown in Table~\ref{table:sensFull}, a 20\% increase in demand turns the availability of supplies into an active constraint. However, counterintuitively, it increases infections marginally. We investigated the reason for this observation, and it was found in the equity objective weight in the master problem ($\lambda_{2,1}$). Under high demand, the equity should be further enforced to prevent states with higher infections and high demand take all the vaccines and leading to myopic decisions. Increasing the state lower bound weights to $\frac{3}{250}$ will reduce the infections to the baseline. This emphasizes the importance of early vaccination in the effectiveness of the process. Since the supply is limited in initial periods, it is the binding constraint, and the extra demand will be satisfied through the supply available at the end of the problem horizon. Although satisfying the remaining demand is important in the long run, it does not reduce the infections within our time horizon, given the delay in the effectiveness of vaccines.
In contrast, a 20\% decrease in demand can result in over one million additional infections in six months. In summary, changes in supply and demand depend on shifts in the binding constraint; increasing demand activates the supply availability constraint, and reducing demand deactivates it. To enhance the interpretability of the sensitivity analysis results, Figure~\ref{appendix-fig:sensitivity analysis} in Appendix~\ref{appendix: Appendix 6-3} provides a graphical summary of the results.
 \looseness-1


The sensitivity analysis reveals the model's sensitivity to changes in the infection rate, followed by the effectiveness and budget of the vaccine. Given the limits on budget and supply, capacity changes have minimal impact, indicating that the system is not capacity-bounded under baseline conditions. Demand elasticity also influences outcomes: when fewer people seek vaccination, cases increase even if vaccine supply and infrastructure remain constant. This highlights that sufficient resources alone are not enough; public willingness and ability to access vaccination are equally critical to achieving effective epidemic control. Overall, the system behaves non-linearly: increases in supply, budget, or infrastructure do not always yield proportionate health benefits. Strategic, data-driven allocation policies tailored to the dynamics of each scenario are necessary for efficient allocation of resources, which underscores the need for efficient solution methods. In addition to these findings, we investigated the effect of vaccination timing. The results show that delay in the vaccination campaign by one month would have resulted in more $900k$ additional infections. This further reinforces the value of early intervention in the response to epidemics and the critical role of timely vaccination deployment in reducing case numbers, supported by the previous literature \citep{Yin2024covid}.\looseness-1


\section{Discussion}\label{discussion}
In this paper, we develop a data-driven optimization framework that integrates epidemic dynamics with efficient and fair resource allocation. Specifically, we propose and validate a need-based vaccine allocation approach, demonstrating its potential for improved pandemic management. To this end, we introduce a novel equity definition within a knapsack problem structure with empirically derived coefficients. Numerical experiments illustrate that our proposed equity criterion substantially outperforms the commonly used Gini-based definition in both efficiency and effectiveness.\looseness-1

Our comprehensive COVID-19 case study for the U.S. highlights the practical viability of the model. To manage computational complexity in large-scale scenarios, we utilize two strategies: (i) simplifying the epidemiological model by using an effective infection rate, thus reducing compartmental complexity, and (ii) designing two heuristic algorithms that deliver high-quality solutions several thousand times faster than the full-scale optimization. These efficiency gains are critical for large-scale stochastic optimization scenarios, allowing practical decision-making without cumbersome computational demands.\looseness-1

Applying our framework retrospectively to the COVID-19 pandemic scenario in the United States, we estimate that a more effective vaccine allocation could have prevented nearly two million infections over a six-month period and potentially saved more than $30,000$ lives, considering the high mortality rate of COVID-19 in 2021 \citep{owid-mortality-risk-covid}. The benefits of our approach become even more pronounced in long-term implementation scenarios, where epidemiological dynamics, vaccine-induced immunity, and access to care are explicitly accounted for allocation strategies. Although demonstrated using COVID-19, our framework is sufficiently generalizable to adapt to other epidemic contexts, requiring minimal modifications in the compartmental and logistical components of the model. \looseness-1


A crucial policy implication of our model is related to the interplay between equity and efficiency. Traditionally, equity in resource allocation is viewed as inherently opposed to efficiency. However, our results of single-period decomposition highlight how equity considerations significantly improve the epidemiological effectiveness of reactive policies during temporally decomposed decision-making. This observation demonstrates that a balanced and equitable allocation strategy can yield enhanced long-term efficiency, particularly with competing regional vaccine demands. This result aligns with previous research on efficiency-equity trade-offs in public health resource allocation \citep{le1990equity,islam2022modeling}. Under limited supply and high infection rates, equitable allocation approaches naturally converge toward equal per-capita distribution, underscoring the value of proactive and sustained vaccination rather than reactive measures. Although prioritizing immediate vaccine allocation to the region with the highest infections may seem beneficial in the short term, our analysis indicates that such myopic policies compromise long-term protection throughout the broader population, potentially amplifying future transmission risks \citep{long2018spatial}. Specifically, prioritizing vulnerable, high-risk populations, including essential workers in high-contact occupations who cannot self-isolate, via an SVI index used to weigh the model priorities, reduces immediate disease burden and diminishes future transmission risks. Hence, equity integration transforms reactive, single-period interventions into strategies that better approximate the proactive, long-term benefits of a multi-period model.\looseness-1

From a methodological viewpoint, our multi-period framework provides an upper-bound benchmark, emphasizing the importance of holistic, forward-looking decision-making rather than isolated myopic decisions. Although heuristic budget adjustments offer efficient approximations, complete multi-period optimization remains crucial for avoiding suboptimal short-term solutions. Given real-world constraints, our decomposition approach provides intuitive guidance for the design of vaccination policies over intermediate horizons, particularly short of achieving herd immunity. \looseness-1

An important practical aspect highlighted by our framework is the prudent use of economic and human resources, which often pose active constraints during epidemic responses. Through a carefully constructed knapsack-based equity formulation, our approach efficiently combines equitable vaccine access with operational effectiveness, enabling significant reductions in mortality without additional economic burden. In addition, our framework supports strategic decisions about the establishment of mass vaccination centers, a critical intervention for rapid, large-scale immunization.\looseness-1

Although effective in accelerating vaccinations, mass vaccination centers involve substantial costs and logistical complexity, including the risk of disease transmission at crowded sites, a potential area for future research. Our model systematically informs these large-scale logistic investments considering both demand and operational necessity. Contrary to traditional approaches that advocate early mass center establishment, our analysis demonstrates that, in resource-rich contexts (such as the US), initial vaccine availability typically restricts rollout more than infrastructure capacity. Consequently, our model initially prioritizes efficient vaccine distribution through existing infrastructure (e.g., pharmacies), transitioning to mass vaccination centers only when necessary. This dynamic allocation facilitates resource-efficient deployment, optimal utilization of the available workforce, and ultimately improves overall vaccination effectiveness\looseness-1.

Incorporating behavioral and economic parameters into the proposed allocation framework requires access to real-time data. Despite challenges in estimating these parameters at the beginning of the vaccination process, frequent screening and data acquisition can be extremely helpful \citep{rabil2022effective}. Regardless, we use SARIMA analysis to validate the generalizability of our model and its projection to the future, in the absence of infection rate data. This framework is of high value in less-advantaged countries where access to population data is limited. By integrating the data in our model as well as dynamic need-based allocations, especially in the initial months of vaccination, rather than population-based allocations, vaccine wastage and delays in the vaccination process are minimized. According to our numerical experiments, demand seems to be a binding constraint in certain areas despite the abundance of vaccine supply. Given the importance of the demand for the effectiveness of the vaccination process, policies such as vaccine education and incentives are recommended to improve public perception and combat vaccine hesitancy \citep{Serra-Garcia20231037}. \looseness-1

 As some recommendations for future researchers, first, the parameter estimation can be extended to scenarios without data available, using tools such as Neural Networks~\citep{niazkar2020application}. Secondly, although our one-dose vaccination framework significantly reduces infections, a two-dose model is expected to be more effective. In an extended model with two-dose vaccines, separate compartments could be defined, with a lower effective infection rate. The deceased compartment also must be defined and adjusted, given the impact of complete vaccination on the mortality rate~\citep{nordstrom2022risk, mcmenamin2022vaccine}. In addition, incorporating other interventions, such as treatment, is a promising direction for future research. This modification would increase the complexity of estimating the effective infection rate; however, thanks to the integrated structure and efficient solution methodology developed in this paper, this complexity can be tackled. Moreover, the issue of jurisdictional authority in vaccine allocation can be examined at a more granular level. In addition to a centralized decision-maker who determines allocation strategies, the model can be extended to include regional and sub-regional authorities' willingness to utilize model recommendations. This extension can offer insight into decentralized decision making for public health interventions.\looseness-1

In conclusion, our findings underscore the critical importance of early investments in robust data collection infrastructure to enable informed, equitable, and effective resource allocation during epidemics. We show that rolling allocation systems, those that incorporate real-time updates on demand and adapt dynamically to evolving epidemic conditions, are not only beneficial but essential for responsive and resilient public health strategies. Establishing transparent, data-driven prioritization criteria further improves both the equity and efficiency of vaccine distribution. More broadly, our integrated modeling framework offers a practical roadmap for aligning logistic efficiency with population-level health outcomes in dynamic and uncertain epidemic environments.\looseness-1
{We would like to express our sincere gratitude to [acknowledge individuals, organizations, or institutions] for their invaluable contributions to this research. We are also grateful to [mention any additional acknowledgements, such as technical assistance, data providers, or colleagues] for their support and assistance throughout the course of this work.}

\appendix
In this document, we provide the supporting material for our paper. \nameref{appendix: Appendix 0} provides an overview of the notations used in the paper, with the mathematical model presented in the paper. \nameref{appendix: Appendix 1} discusses the structure of our vaccine supply chain. \nameref{appendix: Appendix 2} presents a detailed analysis of our data processing and calibration process. \nameref{appendix: Appendix 3} is allocated for a more detailed discussion of our decomposition heuristic methods. \nameref{appendix: Appendix 4} elaborates on the validation method we use in the paper. Our last appendix, \nameref{appendix: Appendix 6} delivers some further discussion on our results. The references used in this supporting material are appended at the end of the document.



\noindent In this document, we provide the supporting material for our paper. Appendix~\ref{appendix: Appendix 0} provides an overview of the notations used in the paper. Appendix~\ref{appendix: Appendix 1} discusses the structure of our vaccine supply chain. Appendix~\ref{appendix: Appendix 2} presents a detailed analysis of our data processing and calibration process. Appendix~\ref{appendix: Appendix 3} is allocated for a more detailed discussion of our decomposition heuristic methods. Appendix~\ref{appendix: Appendix 4} elaborates on the validation method we use in the paper. Our last appendix, Appendix~\ref{appendix: Appendix 6} provides further discussion of our results. The references used in this supporting material are appended at the end of the document.
\vspace{0.5cm}
\\
\textbf{\LARGE Appendix}

\section{Review of Mathematical Notations}
\label{appendix: Appendix 0}

To accommodate the reader, Appendix~\ref{appendix: Appendix 0} is organized to provide a complete explanation of the notation used in the mathematical formulation. We also present the detailed mathematical model developed in the paper.



\vspace{-0.4cm}

\renewcommand{\theequation}{A1-\arabic{equation}}
\setcounter{equation}{0}
{
\footnotesize\begin{multicols}{2}
\noindent\textbf{Sets and Indices:}
\noindent\begin{itemize}[left=0cm]
\item []$\mathbb{T}$: Set of periods, $\mathbb{T}=\lbrace0,...,|T|\rbrace$. 
\looseness-1 \item []$\mathbb{J}$: Set of regions, $\mathbb{J}=\lbrace1,...,|J|\rbrace$.
\looseness-1 \item []$\mathbb{K}(j)$: Set of sub-regions in region $j$, $\mathbb{K}(j)=\lbrace1,..., |K_j|\rbrace$.
\looseness-1 \item []$\mathbb{L}(k_j)$: Set of vaccine administration centers in sub-region $k_j$, primarily consisting of pharmacies but may also include small clinics, retail health centers, and community health sites, $\mathbb{L}(k_j)=\lbrace1,..., |L^k_j|\rbrace$.
\looseness-1 \item []$\mathbb{O}(j)$: Set of possible mass vaccine administration centers in region $j$, such as community centers, stadiums, and other large facilities used in mass vaccination plans, $\mathbb{O}(j)=\lbrace1,..., |O_{j}|\rbrace$.
\looseness-1 \item []$\mathbb{M}$: Set of vaccine suppliers, $\mathbb{M}=\lbrace1,..., |M|\rbrace$.
        \looseness-1 \item []$t$: Index for periods, where $t \in \mathbb{T}$ 
        \looseness-1 \item []$j$: Index for regions, where $j \in \mathbb{J}$   
        \looseness-1 \item []$k_j, m_j, n_j$: Indices for sub-regions in region $j$, where $k_j, m_j, n_j \in \mathbb{K}(j)$
        \looseness-1 \item [] $l^k_j$: Index for vaccine administration centers in sub-region $k_j$, where $l^k_j\in \mathbb{L}(k_j)$
        \looseness-1 \item [] $o_j$: Index for possible mass vaccine center locations in region $j$, where $o_j \in \mathbb{O}(j)$
        \looseness-1 \item []$i$: Index for vaccine suppliers, where $i \in \mathbb{M}$
\end{itemize}
\noindent\textbf{Epidemic Variables:}
\noindent\begin{itemize}[left=0cm]
      \item []{$S_{j}^{t}$: Susceptible population at period $t$ in region $j$ \phantomsection\label{def:epivar1}}
     {\looseness-1 \item []$I_{j}^{t}$: Infected population at period $t$ in region $j$ including both symptomatic and asymptomatic population\phantomsection\label{def:epivar2}}
     \looseness-1 \item []$V_{j}^{t}$: Vaccinated population at period $t$ in region $j$ who gained full or partial immunity against the disease of interest\looseness-1
      \item []$R_{j}^{t}$: Removed (recovered or deceased) population at period $t$ in region $j$  
     \looseness-1 \item []$\tilde{I}_{j}^{t}$: Population susceptible to reinfection at period $t$ in region $j$ (population who recovered from the disease or are vaccinated but not infected yet, as mentioned in the assumptions)
    \looseness-1 \item [] $ \tau_j^t $: Infection threshold for region $j$ at period $t$, used to determine the need for center opening (this threshold varies over time and is specified by healthcare providers and policymakers based on population size and other regional characteristics)
    
\end{itemize}

\noindent\textbf{Supply Variables:}
\noindent\begin{itemize}[left=0cm]
      \item []$G^{1,t}_{i,j}$: Supply from the manufacturer $i$ to regional hub $j$ at period $t$ (Integer)
     \looseness-1 \item []$G^{2,t}_{j,k_j}$: Supply from the regional hub $j$ to the sub-regional hub $k_j$ at period $t$ (Integer) 
     \looseness-1 \item []$G^{3,t}_{k_j,l^k_j}$: Supply from the sub-regional hub $k_j$ to the local vaccine center $l^k_j$ at period $t$ (Integer)\looseness-1
     \looseness-1 \item []$W^{1,t}_{j}$: Inventory level at regional hub $j$ at period $t$  (Integer)
     \looseness-1 \item []$W^{2,t}_{k_j}$: Inventory level at sub-regional hub $k_j$ at period $t$  (Integer)
    \looseness-1 \item []$W^{3,t}_{l^k_j}$: Inventory level at sub-regional local vaccine center $l^k_j$ at period $t$  (Integer)
\end{itemize}

\noindent\textbf{Linking Variables:}
\noindent\begin{itemize}[left=0cm]
      \item []{$\Psi_{j}^{t}$}: Number of vaccines administered to the susceptible population at period $t$ in region $j$ (Integer)\looseness-1
     \looseness-1 \item []${\Xi}_{j}^{t}$: Number of vaccines administered to 
     recovered population at period $t$ in region $j$ (Integer) \looseness-1
    \looseness-1 \item []${\Phi}_{k_j}^{t}$: Number of vaccines administered to the susceptible population at period $t$ in local centers of sub-region $k_j$ (Integer) \looseness-1
    \looseness-1 \item []${\Omega}_{k_j}^{t}$: Number of vaccines administered to recovered population at period $t$ in local centers of sub-region $k_j$ (Integer) \looseness-1

\end{itemize}

\noindent\textbf{Center Opening decision variables:}
\noindent\begin{itemize}[left=0cm]
 \item []$ X_{o_j}^t \in \{0,1\} $: Binary variable indicating whether facility $ o_j $ is open as a vaccine center in region $ j $ at period $ t $, where $ X_{o_j}^t = 1 $ if the facility is open and $ X_{o_j}^t = 0 $ otherwise.
\looseness-1 \item []$ \Upsilon_j^{I,t} \in \{0,1\} $: Binary variable indicating whether an intervention (e.g., opening a mass vaccination center) is required at period $ t $ due to the number of infected cases exceeding the threshold in region $ j $, where $ \Upsilon_j^{I,t} = 1 $ if the threshold is exceeded and $ \Upsilon_j^{I,t} = 0 $ otherwise.
 \looseness-1 \item []$ \Upsilon_j^{D,t} \in \{0,1\} $: Binary variable indicating whether an intervention (e.g., opening a mass vaccination center) is required at time $ t $ due to demand exceeding the current capacity in region $ j $, where $ \Upsilon_j^{D,t} = 1 $ if demand exceeds capacity and $ \Upsilon_j^{D,t} = 0 $ otherwise.
\end{itemize}

\noindent\textbf{Equity Variables:}
\noindent\begin{itemize}[left=0cm]
    \item []$u_{k_j}$: Vaccine per capita in sub-region $k_j$ over the time horizon\looseness-1
    \looseness-1 \item []$v_{m_j,n_j}$: Pairwise absolute-value of difference in vaccine per capita for sub-regions $m_j$ and $n_j$\looseness-1
    \looseness-1 \item []$\bar{u}_{j}$: Average vaccine per capita for region $j$\looseness-1
    \looseness-1 \item []$G(u_j)$: Vaccine Gini coefficient for region $j$, which is defined in equation \eqref{eq4.5} as half of the sum of the absolute pairwise sub-region utility (planned vaccine administered per capita) difference to the average sub-region utility in the region $j$ \looseness-1
    \looseness-1 \item []$\eta$: Maximum of vaccine Gini coefficient among all regions\looseness-1
    \looseness-1 \item []$\zeta_t$: Minimum number of vaccines per capita among all regions allocated to a region at period $t$\looseness-1
\end{itemize}

\noindent\textbf{Parameters:}
\noindent\begin{itemize}[left=0cm]
     \item []{$\mu$}: Recruitment rate (natural death and birth rate) of the population \looseness-1
    \looseness-1 \item []{$\tilde{\beta}_{j}^{t}$}: Transmission coefficient (effective infection rate) between compartments S and I (unvaccinated population) at period $t$ in region $j$\looseness-1
    \looseness-1 \item []$\tilde{\beta}_{1,j}^{t}$: Disease transmission rate in the vaccinated population at period $t$ in region $j$ \looseness-1
    \looseness-1 \item []{$\gamma_1$}: Rate of obtaining immunity in the vaccinated population\phantomsection\label{def:param4}\looseness-1
     \looseness-1 \item []$\gamma$: Removal (recovery or death) rate of infected individuals\looseness-1
    \looseness-1 \item []$\tilde{\sigma}^{t}$: Reinfection rate at period $t$ 
    \looseness-1 \item []$t_r$: Reinfection immunity period\looseness-1
    \looseness-1 \item [] $ \psi \in \{0,1\} $: Binary parameter indicating the existence of long-term immunity provided by vaccination against the disease, where $ \psi = 1 $ if immunity is lifelong or extends beyond the problem's time horizon, and $ \psi = 0 $ otherwise.
     \looseness-1 \item []$N_{k_j}$: Population above 12 years old of sub-region $k_j$ at time 0\looseness-1
     \looseness-1 \item []$N_{j}$: Population above 12 years old of region $j$ at time 0 
     \looseness-1 \item []$c_{o_j}$: Cost of opening center $o_j$ at period $t$ (might vary over time and in different regions) 
     \looseness-1 \item []$c_j^t$: Cost per unit of vaccine administration in region $j$ at period $t$ (might vary over time and in different regions) 
     \looseness-1 \item []$g^{t}_{i}$: Cost of vaccine per dose from manufacturer $i$ at period $t$\looseness-1
     \looseness-1 \item []$g^{1,t}_{i,j}$: Cost per unit of transportation from manufacturer $i$ to regional hub $j$ at period $t$\looseness-1
     \looseness-1 \item []$g^{2,t}_{j,k_j}$: Cost per unit of transportation from regional hub $j$ to sub-region hub $k_j$ at period $t$\looseness-1
     \looseness-1 \item []$g^{3,t}_{k_j,l_j^k}$: Cost per unit of transportation from sub-regional hub $k_j$ to vaccine centers $l_j^k$ at period $t$\looseness-1
     \looseness-1 \item []$w^{1,t}_{j}$: Cost per unit of inventory at regional hub $j$ at period $t$\looseness-1
     \looseness-1 \item []$w^{2,t}_{k_j}$: Cost per unit of inventory at sub-regional hub $k_j$ at period $t$\looseness-1
     \looseness-1 \item []$w^{3,t}_{l_j^k}$:Cost per unit of inventory at vaccine center $l_j^k$ at period $t$\looseness-1
     \looseness-1 \item []$B$: Available national budget
    \looseness-1 \item []$\kappa_{o_j}^t$: The available capacity of mass vaccine center $o_j$ at period $t$
    \looseness-1 \item []$l^0$: The time needed to establish a mass vaccine center
    \looseness-1 \item []$l^1_{i,j}$: The lead time of distribution from supplier $i$ to regional hub $j$
    \looseness-1 \item []$l^2_{j,k_j}$: The lead time of distribution from regional hub $j$ to sub-regional hub $k_j$
    \looseness-1 \item []$l^3_{k_j,l^k_j}$: The lead time of distribution from sub-regional hub $k_j$ to local center $l_j^k$
    \looseness-1 \item []$\chi^t_{k_j}$: The available capacity of local vaccine administration infrastructure (staff, space, etc.) at each period $t$ as a percentage of the population of sub-region $k_j$
     \looseness-1 \item []$\Pi_i^t$: The available capacity of supplier $i$ at period $t$
     \looseness-1 \item []$\xi$: Vaccine wastage percentage (obtained from historical data)
     \looseness-1 \item []$D_j^t$: Expected vaccine demand at period $t$ in region $j$ 
     \looseness-1 \item []$I_j^0$: Total infected population until the beginning of the problem period $0$ in region $j$ 
    \looseness-1 \item []$\tilde{I}_j^0$: Total population susceptible to reinfection until the beginning of the problem period $0$ in region $j$ 
    \looseness-1 \item []$R_j^0$: Total removed population until the beginning of the problem period $0$ in region $j$ 
\end{itemize}
\end{multicols}}

\section{Supply Chain Configuration}
 \label{appendix: Appendix 1}

\begin{wrapfigure}{r}{0.45\textwidth}
    \includegraphics[width=1\linewidth]{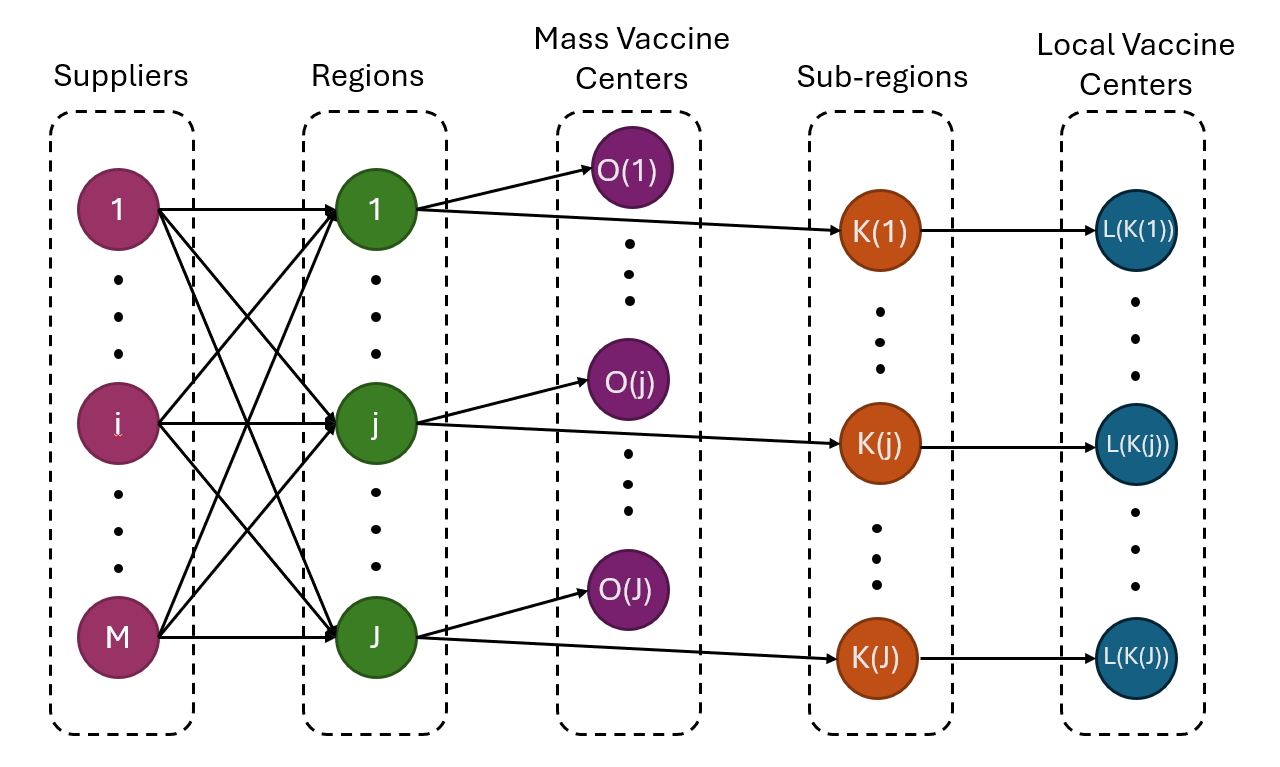}
    \caption{Multi-layer Supply Chain Model}
    \label{appendix-fig:supplychain}
\end{wrapfigure}

\noindent We assume a multi-layer supply chain from suppliers to regions to sub-regions. Figure \ref{appendix-fig:supplychain} provides our traditional multilayer vaccine supply chain. This figure shows the flow of the vaccines from the suppliers to the regions (states) to the mass vaccine centers and sub-regions. Sub-regions allocate the vaccines to the local pharmacies. 
Each supplier can provide the supply to different regions, but each regions can only allocate to its own sub-regions and mass vaccine centers. A similar logic exists within the sub-regional allocations to local pharmacies. 
The multilayer structure presented in Figure \ref{appendix-fig:supplychain} can also be replaced by other models, such as direct-delivery models, where vaccines are shipped directly to administration centers. The choice of a multilayer supply chain is intentional in allowing regional decision makers to allocate vaccines according to their needs. \looseness-1
In the distribution of the COVID-19 vaccine in the US, the vaccines were delivered directly to the vaccination points without passing through regional hubs (\cite{Pfizer2}). This corresponds to a simplified version of our model, which can easily accommodate this setup by removing the hub layer. Another difference is that, while real-world vaccine deliveries often followed a just-in-time approach, our case study assumes biweekly shipments. This difference can be addressed by adjusting the model time periods to align with actual supply chain practices.\looseness-1

\section{Data Analysis and Parameter Calibration} \label{appendix: Appendix 2}
\renewcommand{\theequation}{A3-\arabic{equation}}
\setcounter{equation}{0}

\noindent In this appendix, we briefly provide an overview of our parameter calibration and validation data analysis process. The first step is to recognize the data needs based on the model. We need data to calibrate the model parameters as well as to validate and compare the optimized values of the variables with respect to their actual values.
We gather time-series data for the number of cases and vaccines administered per county and state. Each state has its own abbreviation as its unique ID, and each county has its Federal Information Processing Standard (FIPS) code as its distinct ID. Each date, whether daily, weekly, or biweekly, is also a temporal ID for the data. Within different data sets, we use these unique keys to identify and aggregate the data. \looseness-1

The process of data pre-processing can be summarized as follows: data acquisition and integration, handling missing data, handling data discrepancies, aggregation, and extraction. For data series ID integration, we make sure to aggregate all the data based on the maximum period. This is because we integrate multiple datasets, each with different time scales. We identify each data point by the time stamp, FIPS, and state abbreviations. We use these IDs to recognize the unique values in the data. We determine unique and equal IDs and resolve the inconsistencies.
For counties with different FIPS, including Bristol Bay, plus Lake and Peninsula, and Yakutat plus Hoonah-Angoon in Alaska and New York City counties, we use aggregation and re-labeling to fill them. \looseness-1

To handle missing values, if the data were missing for more than fifty percent of the periods for each county, we consulted other data sources. This is the case for Hawaii and Texas. For other missing values, we use linear extrapolation and interpolation aligned with logical inference (to ensure consistency) to fill the missing values. The missing values are for attributes population over 12 and the number of vaccines administered. For counties marked as ``unknown,'' we filled the value of the missing data with the last available data for the unknown county in that state. States such as AL, VA, MA, NE, and NM have data inconsistency issues, which are handled using interpolation of the consistent data.
Please refer to the section titled ``Case Study Data`` in the original manuscript for the calculation of the infection rate parameters. A more comprehensive list of data sources used for the calibration of our parameters can be found in Table \ref{appendix-DataTable}.\looseness-1

Since data on the storage and transportation costs of COVID-19 vaccines are not abundant, we use the estimates used by \cite{Yin2024covid}, and similarly to them, we consider the variable cost of shipment and storage rather than their fixed cost. Since our main focus is on equitable and infection-averse performance of the supply chain, we use some basic estimates for the cost parameters. To find the cost of vaccine transportation, we need to find the location of the distributors and the modes of transportation. Based on the information captured on the Pfizer website, the main distribution locations are Pleasant Prairie, WI, and Kalamazoo, MI (\cite{Pfizer}). For Moderna, the distribution location was mainly Portsmouth, New Hampshire (\cite{Moderna1}). According to their data, their main distributor is McKesson, which is a centralized distributor. We used the sources mentioned above to estimate the value of$c_j^t$, $g^{t}_{i}$, $g^{1,t}_{i,j}$, $g^{2,t}_{j,k_j}$, $g^{3,t}_{k_j,l_j^k}$, $w^{1,t}_{j}$, $w^{2,t}_{k_j}$, $w^{3,t}_{l_j^k}$, and $B$. The parameters $l^1_{i,j}$, $l^2_{j,k_j}$, and $l^3_{k_j,l^k_j}$ are assumed to be zero given the just-in-time distribution policies, which is a justified assumption since it is always less than each period of the problem (2 weeks) (\cite{Pfizer2}). The duration of the mass vaccination center opening process ($l^0$) is set to 1 period. The average capacity of the mass vaccination centers ($\kappa_{o_j}^t$) is set to 10,000 per period (\cite{le2024mass}). In addition, the parameter $ \psi \in \{0,1\} $, which shows the long-term immunity caused by the COVID-19 vaccine, is zero (\cite{effecvacdur}).\looseness-1

\begin{table}[h]
{\footnotesize
  \centering
  \caption{Data Acquisition Sources }\label{appendix-DataTable}
    \begin{tabular}{lp{10.3cm}}\hline\hline
    Data & Source\\
    \hline
    Infection Data& \cite{CDCState}, \cite{NYTimes}\\
    State and County Vaccination& \cite{CDC}, \cite{Hawaii}, \cite{TexasData0}
    \\
    Vaccine Hesitancy & \cite{DataHesitancy}, \cite{PAUL2022100317}\\
    Population& \cite{CDCState}, \cite{CDC}\\
    Number of Pharmacies& \cite{CensusPharma}\\
    Vaccination Data& \cite{CDC}, \cite{VacSupplyKFF}, \cite{VacSupplyWarr}, \cite{Moderna1}, \cite{Pfizer}, \cite{Yin2024covid}, \cite{effecvacdur},\cite{Colorado},\cite{NHData},\cite{TexasData0}\\
    Vaccine Supply & \cite{VacSupplyKFF}, \cite{CDCJusrisPfizer},\cite{CDCJusrisModerna}, \cite{CDCJusrisJan}\\
    Equity Data& \cite{CDC}, \cite{healthaccess}, \cite{svicounty}\\
    Unreported Cases&\cite{Harvard}, \cite{NIHUnreported}, \cite{CIDRAPUnreported}, \cite{kalish2021undiagnosed}\\
    Reinfection&\cite{chen2024does}, \cite{shastri2021severe}\\
    Recovery and Natural Death and Birth Rate&\cite{liu2008svir}, \cite{Wiersinga2020782}\\
    County Capacity and Distance&\cite{CensusPharma},\cite{CountyDistance}\\
    \hline
    \end{tabular}%
    }
\end{table}%

\subsection{Knapsack Weights}\label{appendix: Appendix 2-1}
\begin{wrapfigure}{r}{0.3\textwidth}
\vspace{-0.2cm}
       \includegraphics[width=\linewidth]{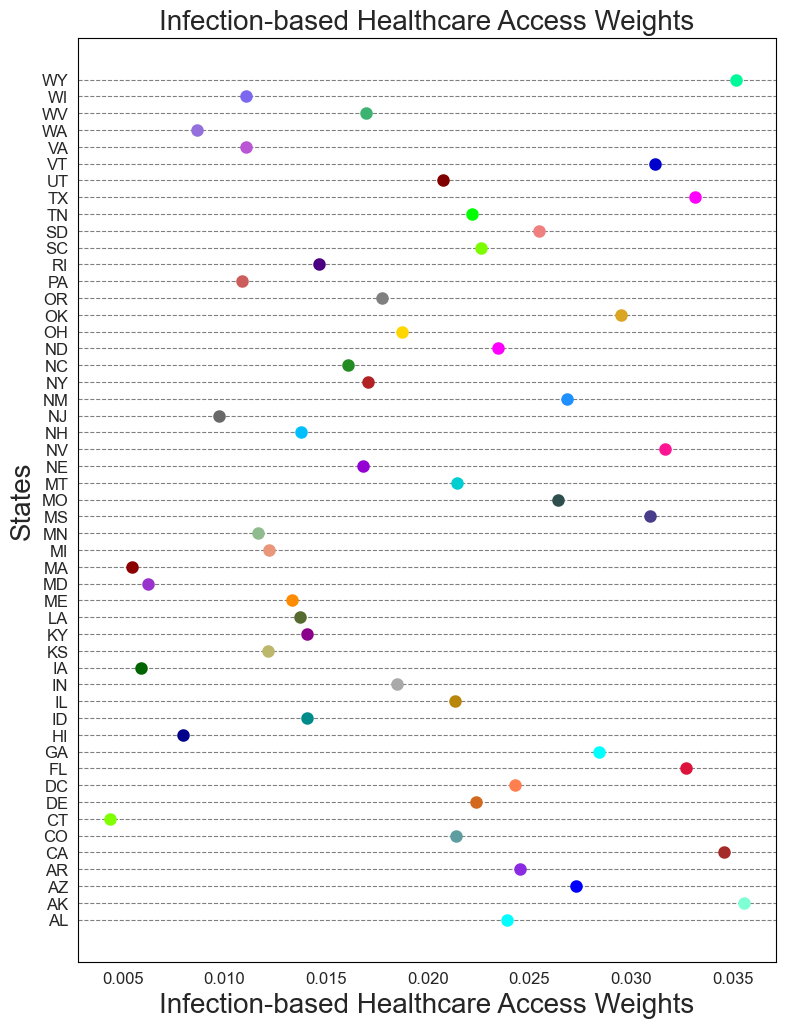}
   \caption{ \footnotesize \centering Healthcare access index in 50 states of the U.S. and DC}
   \label{fig: state_knapsack}
   \vspace{-0.8cm}
\end{wrapfigure}

\noindent As previously described in \nameref{model}, we formulate a knapsack-like equity objective function in Equation~\eqref{objective1-eq1.1}. This function incorporates two key terms: \( (1 - A_j) \) and \( \rho_{k_j} \). The term \( A_j \) represents the level of access to healthcare in each state, calculated using the ranking data from \cite{healthaccess}. To smooth these rankings and reduce skewness, we apply a logarithmic transformation. We then combine these adjusted access scores with state-level population data (\cite{CDCState}) and infection rates to derive a relative weight for each state. These weights are assembled into state-level vectors from which we compute the Euclidean distances. To normalize the scale and mitigate the numerical instability caused by small coefficients, we divide the resulting values by their total sum. The final infection-based healthcare access weights for each state are visualized in Figure~\ref{fig: state_knapsack}.
The second term, \( \rho_{k_j} \), reflects county-level equity considerations and was introduced earlier. It is computed using Social Vulnerability Index (SVI), infection rates, and population data obtained from \cite{CDC}. Specifically, we define \( \rho_{k_j} = \frac{d_{k_j}}{\sum_{k_j} d_{k_j}} \), where \( d_{k_j} = \sqrt{(d^{\text{svi}}_{k_j})^2 + (d^{\beta}_{k_j})^2 + (d^p_{k_j})^2} \). $d^{svi}_{k_j}$ is the relative discretized SVI, calculated for each county by discretizing the SVI into four categories and dividing it by the maximum of these discretized values for the counties in the state to which the county belongs. $d^{svi}_{k_j}$ of county $k_j$ computed by dividing the SVI of county $k_j$ by the maximum of SVI in in counties of state $j$, to which the county belongs. 
$d^{\beta}_{k_j}$ is calculated by dividing the effective infection rate of the county $k_j$ by the maximum effective infection rate in the counties of state $j$. Similarly, $d^{p}_{k_j}$ is yielded by dividing the population of county $k_j$ by the maximum of the population in counties of state $j$. This composite distance metric captures the relative vulnerability of each county. This formulation allows the model to allocate resources proportionally based on normalized need across counties within a state.
The log-transformed values of these equitable knapsack weights are presented in Figure~\ref{fig: county_knapsack} in the main paper. The density and spread within each state reveal significant intra-state disparities, reinforcing the need for granular (county-level) rather than coarse (state-level) allocation mechanisms. States such as Texas, California, Florida, and New York exhibit particularly wide spreads and dense distributions, likely reflecting their larger populations and greater internal heterogeneity. 
\textbf{Figure~\ref{fig: state_knapsack}} illustrates the variation in infection-based healthcare access weights across U.S. states, highlighting disparities in access and epidemiological burden. \looseness-1






\subsection{Removal (Death and Recovery) Rate}\label{appendix: Appendix 2-2}

\noindent For calculating parameters such as $\gamma$, $\mu$, and $\gamma_1$ we use the methodology suggested by \cite{liu2008svir}. Based on this methodology, the natural Death and Birth Rate ($\mu$) is calculated as the ratio of the number of days in each period (14 days) to the life expectancy in days ($\sim 75$ years). For $\gamma$ and $\gamma_1$, we use the existing literature, which may not be accurate, and therefore calibrate the parameters in the model for better accuracy. We set the upper bounds of the vaccine variables at the actual values with budget and supply availability constraints, without any equity consideration. We then change the values of the parameters within the range suggested by the literature until we get close enough to the actual infected cases. We also use \cite {Blanchini2023} to validate our calculated removal rate and other temporal rates.

\subsection{Vaccine Effectiveness}\label{appendix: Appendix 2-3}

\noindent Since our model focuses on the initial phases of vaccination and according to \cite{pilishvili2021interim}, and the effectiveness of one vaccine dose and two vaccine doses are 82 and 94 percent, respectively, it is a justified assumption to limit our focus to only one dose for this article to address the integration of epidemic and supply chain modeling rather than the complexity of the model.
  \cite{abu2021pfizer} has another estimate of the effectiveness of a one-dose vaccine against infection, which is about 50 to 70 percent. The difference between the effectiveness of one and two doses of vaccine is between 10 and 30 percent, according to the reviewed literature. However, if we consider the two-month required period between the administration of the first and second doses of the vaccine with effectiveness delay, only vaccinated people within the initial six periods of the entire twelve periods of our model will be affected by the effects of the second dose of the vaccine. The total of second doses administered is 40 million according to existing data from \cite{CDC}. This amount is insignificant given the 10 to 30 percent difference in effectiveness of vaccination in one- and two-dose vaccines. For the final calibration, we implemented a similar approach in the previous section to achieve the most suitable parameter within the range suggested by the literature. In detail, we rewrite the infection term, which was previously $\big( \tilde{\beta}_{j}^t S^{t}_{j} I^{t}_{j} +\tilde{\beta}_{j}^{1,t} V^{t}_{j} I^{t}_{j} \big)$ \eqref{objective1-eq1.1}, as $\big( \tilde{\beta}_{j}^t S^{t}_{j} I^{t}_{j} +\tilde{\beta}_{j}^{1,t} V^{1,t}_{j} I^{t}_{j}+\tilde{\beta}_{j}^{2,t} V^{2,t}_{j} I^{t}_{j} \big)$. The variable $V^{2,t}_{j}$ is the second dose compartment and $\tilde{\beta}_{j}^{2,t}$ is the effectiveness of two doses of vaccines. Since there is only inflow from the first dose vaccine compartment to the second dose compartment, we expect that at some rate the population in the first dose compartment will decrease and move to the second dose vaccine compartment (the rate would be the rate of getting the second dose of the vaccine). For instance, if ten percent of the population gets the second dose of the vaccine, and the infection rate in this population is almost 80 percent of the first-dose compartment, assuming the most significant flows are infection and vaccination flows, we have roughly $0.9\tilde{\beta}_{j}^{1,t}V^{1,t}_{j} I^{t}_{j}$ as the first dose infections and $(0.8)\tilde{\beta}_{j}^{1,t}(0.1V^{1,t}_{j}+V^{2,t}_{j}) I^{t}_{j}$ as the second dose infections. So, if we start with $V^{2,0}_{j}=0$, we will have the total inflow of $\sum_{i=0}^t 0.1iV^{1,i}_{j}$ at time $t$. Therefore, the upper limit of infections will be less than one dose of vaccine, given the decrease in the infection rate. This will lead to an overestimation of the infection rate in our vaccinated population in our calibrated model. 

 Since our model focuses on the initial phases of vaccination, and according to \cite{pilishvili2021interim}, the effectiveness of one and two vaccine doses is 82\% and 94\%, respectively, {it is reasonable to focus solely on one-dose vaccination in this study. This allows us to integrate epidemic and supply chain modeling without added model complexity.}

\subsection{Vaccination Rate}\label{appendix: Appendix 2-4} 

\noindent We use vaccination data from \cite{CDC}, which provides county-level records for all 50 U.S. states and DC. {However, data for Texas and Hawaii were missing for our time frame, so we supplemented them using sources from \cite{Hawaii} and \cite{TexasData0}.} In some states, only vaccine administration percentages were available. {We estimated missing counts by multiplying the administration percentage by the population over age 12 at each time point.} The vaccine data from CDC had inconsistency with other sources for Colorado and New Hampshire; therefore, we consulted the state health department data for these two states \citep{Colorado,NHData}. {Our model does not directly define a vaccination rate; instead, it includes variables for the number of vaccines administered \((\Psi, \Xi, \Phi, \Omega)\).} We impose upper bounds on these variables based on demand and capacity. {Vaccine willingness estimates from \cite{DataHesitancy} were used to approximate county-level demand, which we then regionally calibrated by adjusting a scaling factor so that model-generated vaccine administration matches observed data.} {During calibration, we initially set county-level capacity sufficiently high to ensure that demand, not supply, was the limiting factor, and vice versa.} {We used \cite{CensusPharma} to estimate pharmacy-level administration capacity based on pharmacy size and staff, and adjusted final rates to match actual total vaccine distribution.}\looseness-1

\subsection{Summary of Model Input Data}\label{appendix: Appendix 2-5} 

\noindent Table \ref{appendix-table: param_summary_ref} provides a summary of the values of the most important parameters in the model. The first column provides the name and notation of the parameter consistent with the notation used. The second column provides the calibrated value of the parameters or, alternatively, a range for them in spatially/temporally varying parameters. The third column presents the references we use to calibrate the model parameters.

{\scriptsize
\begin{longtblr}[
    caption={Model Parameters Calibrated Values}, 
    label={appendix-table: param_summary_ref}]{
  colspec={>{\arraybackslash}p{6cm}
           >{\centering\arraybackslash}p{4cm}
           >{\arraybackslash}p{5cm}}
}
    \hline[2pt]
    Parameter & Calibrated Value& Reference\\
    \hline
    Initial Infected Population ($I_j^0$) & Varies spatiotemporally & \cite{CDCState},\cite{NYTimes} \\
    Natural Death and Birth Rate ($\mu$) & $\frac{14}{75*365}$ &\cite{liu2008svir}\\
    Infection Rate ($\tilde{\beta}_j^t)$& Varies by time and region (Infection Rate)&\cite{liu2008svir}, \cite{Yin2024covid}\\
    Ratio of Vaccine Willingness in Unvaccinated to Vaccinated Population& 1&\cite{PAUL2022100317}\\
    Removal Rate ($\gamma$) & 1 &\cite{Wiersinga2020782}\\  
    Average First-dose Vaccine Effectiveness Rate ($\propto \tilde{\beta}_j^{1,t}$) & 0.8&\cite{Yin2024covid}\\
    Vaccine Effectiveness Period ($\gamma_1$)& 1&\cite{effecvacdur}\\
    Vaccine Wastage Rate ($\xi$) & 0 &\cite{wastageNBC}\\
    Reinfection Rate ($\tilde{\sigma}^t$,$t_r$)& $0-\frac{1}{6}$&\cite{chen2024does}, \cite{shastri2021severe}\\
    Expected Demand ($D_j^t$) & 60\%-98\% of the population& \cite{DataHesitancy}, \cite{CDC} \\
    Pharmacy Data ($l(k_j)$,$\chi^t_{k_j}$)& Varies by the region&\cite{CensusPharma} \\
    Vaccine Cost ($c_{o_j}$) &10-24& \cite{VacSupplyKFF} \\
    Vaccine Supply ($\Pi_i^t$) &Varies by time and region& \cite{CDCJusrisPfizer},\cite{CDCJusrisModerna}, \cite{CDCJusrisJan} \\
    Population Data ($N_{k_j},N_{j}$) & Varies Geographically& \cite{CDC}\\
    Social Vulnerability Index (SVI) & Varies Spatially & \cite{CDC}, \cite{svicounty}\\
    \hline[2pt]
\end{longtblr}}






\subsection{Infection Calibration}\label{appendix: Appendix 2-6}

\noindent As mentioned in the main text, our calibration of the infection rate ($\tilde{\beta}$) is based on a set of equations presented in the matrix \eqref{beta:matrix1}. These equations, \eqref{beta:matrix1}, are derived from the model proposed by \cite{liu2008svir} and are used to account for the calibration requirements described above. {They allow us to tailor the effective infection rate to fit both our model structure and available data.} {The first row of the matrix corresponds to constraint \eqref{eq2.2}, where we use historical data to estimate the future susceptible population.} The second row reflects the vaccinated population and corresponds to constraint \eqref{eq2.4}, {while the third row represents the infection dynamics captured in equation \eqref{eq2.3}.}
{\footnotesize
\begin{flalign}\label{beta:matrix1}
\begin{bmatrix}
    1-\mu-\tilde{\beta}_{j}^t I_{j}^t &0&0&- \Psi_{j}^t\\
    0&1-\mu-\gamma_1-\tilde{\beta}_{j}^{1,t} I_{j}^t&0&\Psi_{j}^t\\
    0&0&1-\mu-\gamma &\tilde{\beta}_{j}^{1,t} V^{t}_{j}I^{t}_{j} +\tilde{\beta}_{j} S^{t}_{j}I^{t}_{j}\\
    \end{bmatrix}
    \begin{bmatrix}
    S^{t}_{j}\\
    V^{t}_{j}\\
    I^{t}_{j}\\
    1
    \end{bmatrix}    
    =
    \begin{bmatrix}
    S^{t+1}_{j}\\
    V^{t+1}_{j}\\
    I^{t+1}_{j}\\
    \end{bmatrix} 
\end{flalign}
}

\vspace{-0.8cm}

\section{Decomposition Algorithms} \label{appendix: Appendix 3}

\noindent In this section, we provide details on the decomposition methods, Gini-based and Knapsack-based, presented to solve the proposed large-scale formulation.

\subsection{Gini-based Decomposition} \label{appendix: Appendix 3-1}

\begin{figure}[h]
\vspace{-0.3cm}
    \centering
    \includegraphics[width=0.77\linewidth]{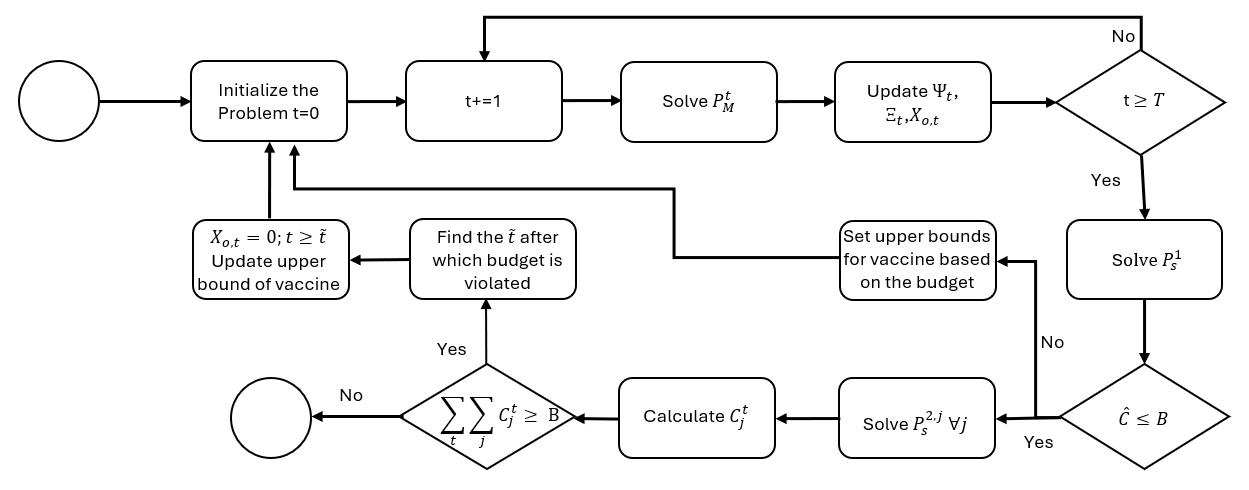}
    \caption{\centering Modified Spatio-temporal Decomposition}
    \label{appendix-fig:decom-sptem}
\end{figure}

\noindent {The Gini-based Decomposition algorithm \eqref{appendix-dec1-objective1-eq1.1}--\eqref{appendix-dec1-eq3} addresses the scalability limitations of the full Gini-based Formulation \eqref{objective1-eq1.1}--\eqref{eq6.10} by decomposing it into a master problem (policy level) and subproblems (operational level).} Initially, the problem was decomposed spatially; initial numerical experiments showed that while the spatial decomposition improves scalability over the full formulation, its efficiency gains are limited. To address this, we implement a modified algorithm—outlined in Figure~\ref{appendix-fig:decom-sptem}—based on a three-level spatio-temporal decomposition.

{This modified approach is structured as a three-level spatiotemporal decomposition, where each level plays a distinct role in balancing epidemic control, supply logistics, and budget feasibility.} {At the top level, we solve a series of temporally decomposed master problems ($P_M^t$), with objective function \eqref{appendix-dec1-objective1-eq1.1} and constraints adapted from the Knapsack-based Decomposition \eqref{dec2-objective1-eq1.1}--\eqref{dec2-objective3-eq2.1.3}. Each of the $t$ problems takes the optimal solution from previous periods as input.} 
{The middle level consists of a single country-level subproblem ($P_S^1$), which does not consider Gini equity. It uses the optimal vaccine allocation vectors ($\Psi^*_U$, $\Xi^*_U$) from the upper level \eqref{appendix-dec1-eq3-1} and solves a simplified model with constant objective \eqref{appendix-dec1-objective3-eq1.1} to estimate total supply chain cost using formula \eqref{eq3.11}. To elaborate, the middle level is one sub-problem for country-level allocations without Gini coefficient consideration($P_S^1$), which takes the optimal variables from the upper level as parameters \eqref{appendix-dec1-eq3-1} ($\Psi^*_U$ and $\Xi^*_U$ are the upper-level optimal solutions for vectors $\Psi$ and $\Xi$) and solves the subregional-level allocation problem with a constant objective value called $c$ \eqref{appendix-dec1-objective3-eq1.1} to come up with an estimate for the cost of the supply chain using the formula provided in equation \eqref{eq3.11}}. {The third level contains one subproblem per region ($P_S^{2,j}$), each solved independently using objective \eqref{appendix-dec1-objective2-eq1.1}. These regional models refine vaccine distribution under equity constraints.} {Feasibility is verified using \eqref{appendix-dec1-eq3}. To ensure optimality, we enforce:}
\(
\frac{(B - C)}{\min_j \sum_t \frac{1}{T}\hat{c}_j^t} \leq 1 \quad \forall j, t
\)
{This guarantees that the remaining budget is insufficient to purchase more than one additional vaccine dose. The bound can be adjusted to reflect the desired trade-off between efficiency and computational effort.}\looseness-1

 

{\scriptsize
\vspace{-0.3cm}
\begin{flalign}
\textit{\textbf{Upper-level Problem at Period $t$ ($ {P}_M^t$)}}\rightarrow &\min   \qquad \lambda_0\sum_j \big( \tilde{\beta}_{j}^t S^{t}_{j} I^{t}_{j}+\tilde{\beta}_{j}^{1,t} V^{t}_{j} I^{t}_{j}\big) -\lambda_{1,1} \zeta_{t} &\label{appendix-dec1-objective1-eq1.1}
\end{flalign}

\vspace{-0.9cm}

\begin{flalign}
&&\textrm{\textbf{s.t.}}   \qquad
\eqref{eq2.2}, \eqref{eq2.3}, \eqref{eq2.4}, \eqref{eq2.5}, \eqref{eq2.6}, \eqref{eq20.0},
\eqref{eq20.1}, \eqref{eq20.2}, \eqref{eq20.3}, \eqref{eq20.4}, \eqref{eq20.5}, \eqref{eq20.6}, \eqref{eq3.12}, \eqref{eq4.0},\eqref{dec1-eq1}, \eqref{decom2-eq2}, \eqref{decom2-eq4}\nonumber
\end{flalign}

\vspace{-0.9cm}

\begin{flalign}
\textit{\textbf{Middle-level Problem ($ {P}_S^1$)}} \rightarrow &\max   \qquad c &&\label{appendix-dec1-objective3-eq1.1}
\end{flalign}

\vspace{-0.9cm}

\begin{flalign}
\vspace{-1cm}
&&\textrm{\textbf{s.t.}} \qquad  
\eqref{eq3.5}, \eqref{eq3.6}, \eqref{eq3.7}, \eqref{eq3.8}, \eqref{eq3.9}, \eqref{eq3.10}, \qquad
\Psi= \Psi^*_U, \Xi= \Xi^*_U \label{appendix-dec1-eq3-1}
\end{flalign}

\vspace{-0.8cm}
 
\begin{flalign}
\textit{\textbf{Lower-level Problem (${P}_S^{2,j}$)}}\rightarrow&\min \qquad  G(u_j)  \qquad &\forall j&\textrm{\textbf{s.t.}} \qquad \vspace{0.5cm} \eqref{eq20.5}, \eqref{eq20.6}, \eqref{eq4.1}, \eqref{eq4.2},  \eqref{eq4.4}, \eqref{eq4.5}, \eqref{eq4.6}\label{appendix-dec1-objective2-eq1.1}
\end{flalign}

\vspace{-0.8cm}

\begin{flalign}
\textit{\textbf{Feasibility Check}}\rightarrow
&   \sum_j C_j  =   \sum_j \sum_t (c_j^t (\Psi_j^t+\Xi_j^t) +\sum_{i}g^{t}_{i} G^{1,t}_{i,j} +\sum_{i}g^{1,t}_{i,j} G^{1,t}_{i,j}+ \sum_{k_j}g^{2,t}_{j,k_j} G^{2,t}_{j,k_j} \label{appendix-dec1-eq3}
& \\&    +\sum_{l_j^k}\sum_{j_k}g^{3,t}_{k_j,l_j^k} G^{3,t}_{k_j,l_j^k} +w^{1,t}_{j} W^{1,t}_{j}+ \sum_{k_j}w^{2,t}_{k_j} W^{2,t}_{k_j}  + \sum_{l_j^k} w^{3,t}_{l_j^k} W^{3,t}_{l_j^k} + \sum_{o_j}  X_{o_j}^{t} c_{o_j}^{t}) \leq B \nonumber&
\end{flalign}
}

\vspace{-0.5cm}

With the Gini-based Decomposition \eqref{appendix-dec1-objective1-eq1.1}--\eqref{appendix-dec1-eq3} explained, we proceed to providing supplementary material on the Knapsack-based Decomposition \eqref{dec2-objective1-eq1.1}--\eqref{dec2-objective3-eq2.1.3}, which was briefly explained in the main paper.

\subsection{ Knapsack-based Decomposition} \label{appendix: Appendix 3-2}  

\begin{figure}[h]
\vspace{-0.2cm}
    \centering
    \includegraphics[width=0.8\linewidth]{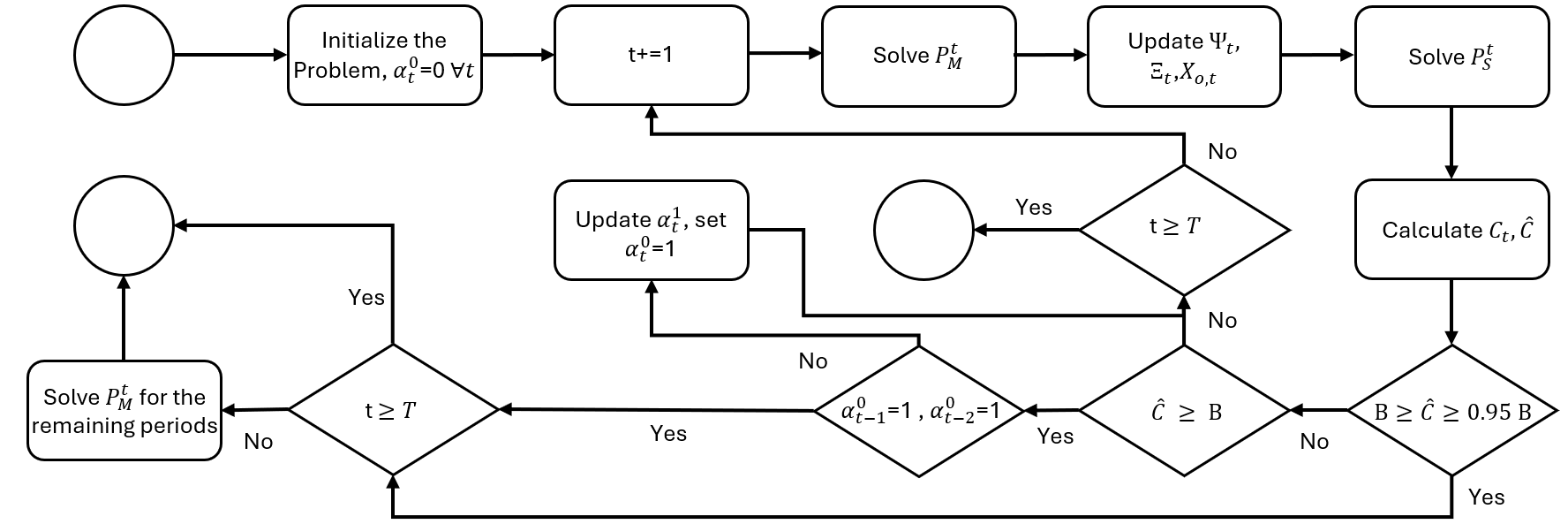}
    \caption{\centering Flow Chart of Knapsack-based Decomposition \eqref{dec2-objective1-eq1.1}--\eqref{dec2-objective3-eq2.1.3}}
    \label{appendix-fig:Knapsack-based Decomposition}
\end{figure}

Figure \ref{appendix-fig:Knapsack-based Decomposition} presents the flowchart of the Knapsack-based Decomposition algorithm \eqref{dec2-objective1-eq1.1}--\eqref{dec2-objective3-eq2.1.3}. The procedure begins with initializing $t = -1$ and setting initial values for all compartmental variables. For each period $t$, we solve the master problem, update the vaccine allocation and center opening decisions at the state level, and then initiate the corresponding subproblem. We calculate the associated supply chain cost, add it to the cumulative cost of previous periods, and compare the total with the available budget. If the cost exceeds 95\% of the budget, we terminate the loop. Based on the current progress through the planning horizon, we either finalize the model or proceed to solve only the master problem in future periods. If the cost is still within budget, we move on to the next iteration. If the cost exceeds the budget, we check whether similar infeasibility has occurred in the previous two periods to avoid infinite cycling. If not, we update the upper bounds on vaccine variables in the master problem ($\alpha^1_t$) and proceed, provided there are remaining periods. If infeasibility has occurred in two consecutive periods, we assume the upper bounds are too loose and either refine them or exit the loop, as illustrated in the figure. The Knapsack-based Decomposition logic follows the same principle as the three-level structure in Figure~\ref{appendix-fig:decom-sptem}, except the latter includes an additional spatial allocation layer at the county level \eqref{appendix-dec1-objective2-eq1.1} to verify budget feasibility after solving both master and subproblems.

In Algorithm~\ref{appendix-alg:two}, we decompose the main problem into a master problem ($P_M^t$) and a sub-problem ($P_S^t$) for each period, and we solve it iteratively. We have an upper bound for our vaccine variables ($\Psi_j^t+\Xi_j^t$), and update this at each iteration based on the vaccine goal and the availability of budget. The logic of this algorithm is based on the behavior of SVIR models in which the earliest possible vaccination is desirable. In this method, we solve the master problem in each period, provide the inputs of it, which are state-level vaccine allocation variables, to the subproblem for the corresponding period, and check the total budget at each iteration until we cover all periods in our time horizon. At each iteration, the supplier availability, the upper bound for the number of vaccines, and the upper bound for the center-opening variables will be updated.

\begin{algorithm}[h]
\caption{Knapsack-based Decomposition \eqref{dec2-objective1-eq1.1}--\eqref{dec2-objective3-eq2.1.3}}\label{appendix-alg:two}
\vspace{.1cm}
{\scriptsize
\KwData{$\mu, \gamma, \gamma_1, \psi, N, \omega, g, g^1, g^2, g^3, w^2, w^3, B, \chi, \tau, \Pi, \upsilon, \tilde{\beta}, \tilde{\beta}_1, \xi$
\hspace{0.1 in}\Comment{\tiny This input will be the output of data pre-processing}} 
\Begin{
$S^0_j$; $I^0_j$; $V^0_j \gets 0$; $R^0_j$; $W_j^{1,0} \gets 0$; $\Psi_j^0 \gets 0$; $\Xi_j^0 \gets 0$; $\hat{C} \gets 0$; $iter \gets 0$; 
$\alpha^1_t \gets 0$; $X_{act} \gets 0$; {$T_1 \gets 0$, $T_2 \gets 1$}; $\forall t \in \mathbb{T}, \forall j\in \mathbb{J}$\;

    \While{$T_2 \leq T$}{
    Solve $P_{M}^t(T_1, T_2, S(T_1), I(T_1), V(T_1), R(T_1), W_j^{1,T_1-1}, \alpha^1_t)$\; 
    Solve $P_{S}^t(T_1, T_2, \Psi_j^{T_1-1}, \Xi_j^{T_1-1}, X_{o_j}^{T_1-1})$\;
    $\hat{C} \gets \hat{C} + \hat{C}_{T_1-1}$\; 
      \If{$0.95 B \leq \hat{C} \leq B$}{$X_{act} \gets 1$\; \textbf{break;}\vspace{-1em}} 
      \If{$\hat{C} > B$}{
          \If {$\alpha^1_{T_1-1}==1$ \& $\alpha^1_{T_1-2}==1$}{ 
              $X_{act} \gets 1$; 
              $\hat{C} \gets \hat{C} -  \hat{C}_{T_1-1}$; 
              $T_1 \gets T_1 - 1$; $T_2 \gets T_2 - 1$\;
              \textbf{break;}\vspace{-1em}
          }\vspace{-1em}
          $\hat{c}_j^{T_1-1} \gets \frac{\sum_i g_i^{T_1} \Pi_i^{T_1-1}}{\sum_i \Pi_i^{T_1-1}} + (g^{2,{T_1-1}}_{j,k_j} + \frac{\sum_{k_j} g^{3,{T_1-1}}_{k_j,l_j^k}}{|K_j|} + w^{1,{T_1-1}}_{j})$\;
          $\sum_j(\Psi_j^{T_1-1} + \Xi_j^{T_1-1}) \leq |J| \frac{(B - \hat{C})}{\sum_j \hat{c}_j^{T_1-1}} \sum_i \Pi_i^{T_1-1}$\; 
          $\hat{C} \gets \hat{C} - \hat{C}_{T_1-1}$; $T_1 \gets T_1 - 1$; $T_2 \gets T_2 - 1$\;\vspace{-1em}
      }
      $T_1 \gets T_1 + 1$; $T_2 \gets T_2 + 1$\; \vspace{-1em}
    }
\If{$X_{act} == 1$}{
    \While{$T_2 \leq T$}{
        $\alpha^1_t \gets 0$\; 
        Solve $P_{M}^t(T_1, T_2, S(T_1), I(T_1), V(T_1), R(T_1), W_j^{1,T_1-1}, \alpha^1_t)$\;
        $T_1 \gets T_1 + 1$; $T_2 \gets T_2 + 1$\;
    }\vspace{-1em}}\vspace{-1em}
}}
\end{algorithm}

Now, we investigate the relationship between Knapsack-based Decomposition \eqref{dec2-objective1-eq1.1}--\eqref{dec2-objective3-eq2.1.3} and Knapsack-based Formulation \eqref{eq2.2}--\eqref{eq7.1} when all objective functions are active. As explained in the main text, we use the scalarization method to combine objective functions. This makes the resulting model very sensitive to the selection of weights. Even if we normalize the objective function, there is still an imbalance between the objectives of Knapsack-based Decomposition \eqref{dec2-objective1-eq1.1}--\eqref{dec2-objective3-eq2.1.3} and Knapsack-based Formulation \eqref{eq2.2}--\eqref{eq7.1}. 
We can see that in each period when the budget is not tight, $P_M^t$ and $P_S$ are relaxations for the original problem. This means that the optimal objective value of the Knapsack-based Decomposition \eqref{dec2-objective1-eq1.1}--\eqref{dec2-objective3-eq2.1.3} is a lower bound for the Knapsack-based Formulation \eqref{eq2.2}--\eqref{eq7.1}. However, when the budget is violated and we add feasibility constraints to the original problem, we overestimate the cost to avoid infeasibility. Therefore, we restrict the relaxation by adding a tighter constraint on the available supply compared to the original problem. This characteristic makes the relationship between Knapsack-based Decomposition \eqref{dec2-objective1-eq1.1}--\eqref{dec2-objective3-eq2.1.3} and Knapsack-based Formulation \eqref{eq2.2}--\eqref{eq7.1} inclusive.

\begin{proposition}\label{appendix-prop1}
The optimal solution produced by the Knapsack-based Decomposition \eqref{dec2-objective1-eq1.1}--\eqref{dec2-objective3-eq2.1.3} is a feasible solution to the Knapsack-based Formulation \eqref{eq2.2}--\eqref{eq7.1}.
\end{proposition}

\proof{Proof of Proposition~\ref{appendix-prop1}} Assume that for each period $t$, optimal solutions to the master problem $P_M^t$ and subproblem $P_S$ exist. We aim to show that these combined solutions satisfy all constraints of the full Knapsack-based Formulation \eqref{eq2.2}--\eqref{eq7.1}. Constraints \eqref{eq2.2}--\eqref{eq4.0} are explicitly included in $P_M^t$ and therefore satisfied by construction for all $j$ and $t$. The restriction \eqref{eq3.4}, which bounds the vaccine variables between zero and the population of each state, is enforced by the definitions of variables in $P_M^t$. To show that constraint \eqref{eq3.6} holds, we observe that constraint \eqref{dec1-eq1} in $P_M^t$ ensures that the sum of regional vaccine allocations does not exceed capacity, while constraint \eqref{eq3.5} in $P_S$ links regional and sub-regional allocations, applying the same capacity bounds to the sub-regional level.\endproof

{Supply flow constraints \eqref{eq3.7}--\eqref{eq3.10} are directly enforced within $P_S$ and thus hold.} {Constraint \eqref{eq3.11}, which ensures budget feasibility, is enforced via feasibility constrain \eqref{decom2-eq4}. These constraints iteratively impose upper bounds on the master problem’s allocations to ensure that the (conservatively overestimated) supply chain cost remains within the budget.} {Therefore, the combined optimal solution from the decomposition satisfies all constraints in the Knapsack-based Formulation, and is thus feasible.}
{ Overall, the proposed decomposition algorithm yields high-quality feasible solutions efficiently.} A potential limitation arises when the budget is extremely tight, requiring trade-offs between proximity to distribution centers and transportation cost. Since the master problem lacks cost information, it may under-prioritize distant but higher-need regions. This issue can be addressed by incorporating a parameter into the master problem that reflects the number of infections averted per unit cost. However, we do not emphasize this limitation, as equity constraints and objective function inherently discourage biased allocation and penalize discriminatory behavior.\looseness-1

\section{Discussion on Calibration and Validation} \label{appendix: Appendix 4}

\noindent As discussed in the main paper, our first validation is done to see if the model's behavior is consistent with reality, when the data are available. Following the discussion in the paper, Figure \ref{vaildtotal} visualizes the model-induced cases along with the actual cases for each period during the 6-month time horizon of the problem. It can be seen that the model reproduces the reality almost perfectly. Figure \ref{vaildstate} shows the linear relationship (slope 1) between the actual and model cases over six months for all states. All p-values are greater than 0.025 for both US-level and state-level validation.
{\centering
\begin{figure}[h]
\vspace{-0.5cm}
\hspace{1em}
\begin{minipage}{.4\textwidth}
    \subfloat{\includegraphics[width=\textwidth]{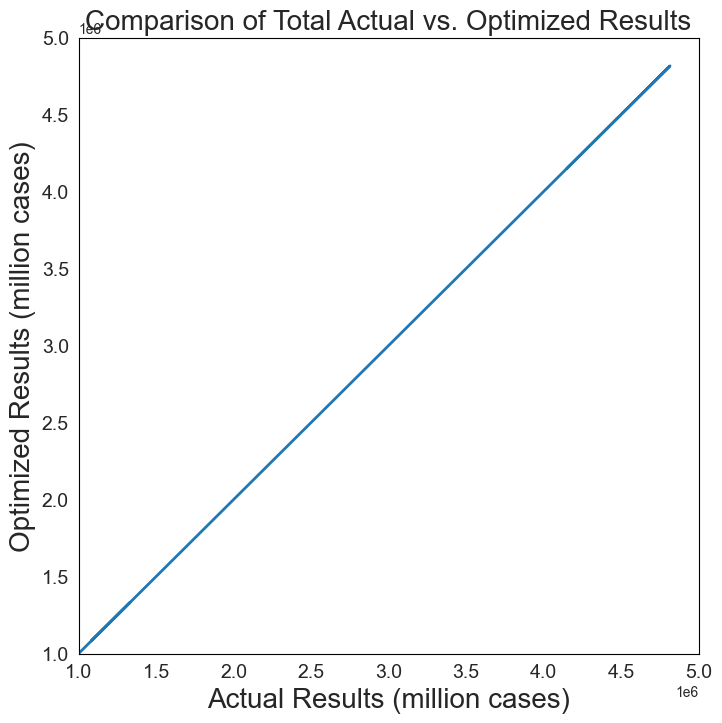}}
       \caption{ \centering {\scriptsize Results of Validation for Total Infection Cases in the U.S. }}\label{vaildtotal}
\end{minipage}  
\hfill
\begin{minipage}{.45\textwidth}
   \subfloat {\includegraphics[width=\textwidth]{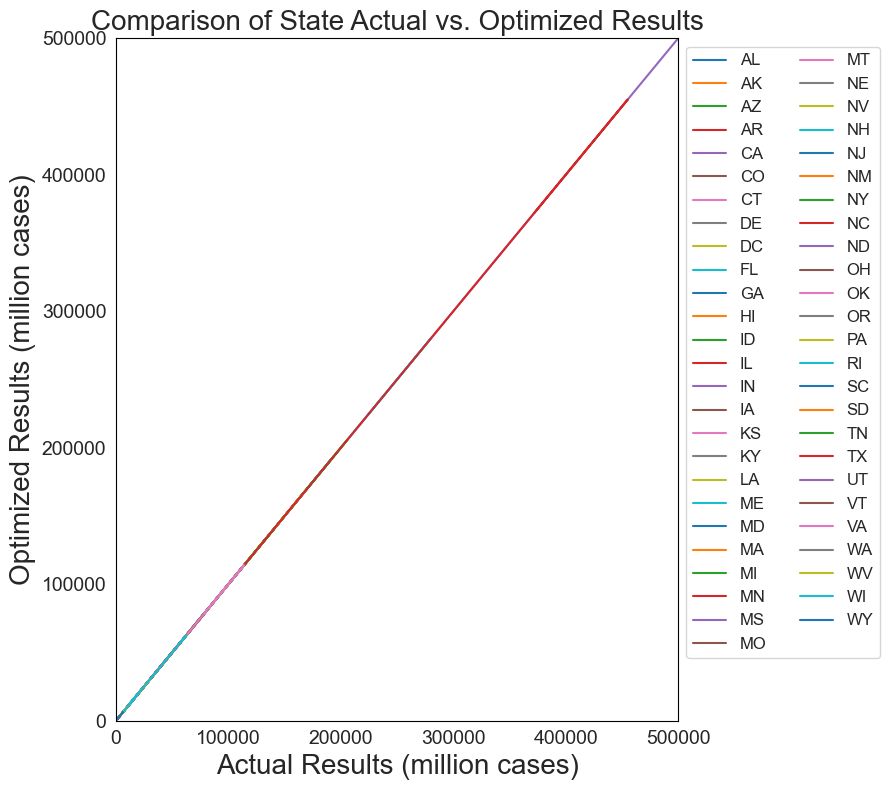}}
   \caption{  \centering {\scriptsize Results of Validation for Total Infection Cases in each State of the U.S.}}\label{vaildstate}
\end{minipage}
\hspace{1em}
\vspace{-0.8cm}
\end{figure}}

\subsection{SARIMA for Validation} \label{appendix: Appendix 4-1}

In this section, we extend the discussion on SARIMA for validation, presented in Section~\ref{sarima} in the main paper. The Seasonal AutoRegressive Integrated Moving Average (SARIMA) model is a generalization of the ARIMA model that supports univariate time series data with a seasonal component. It is denoted by \(
\text{SARIMA}(p, d, q) \times (P, D, Q)_m
\), where $(p, d, q)$ are the non-seasonal components and $(P, D, Q)_m$ are the seasonal components of the model.\(p\) determines the order of the non-seasonal autoregressive (AR) part, \(d\) determines the degree of non-seasonal differencing, and \(q\) determines the order of the non-seasonal moving average (MA) part. On the other hand, \(P\) indicates the order of the seasonal AR part, \(D\) indicates the degree of seasonal differencing, \(Q\) indicates the order of the seasonal MA part, and \(m\) is the length of the seasonal period per unit of time. The model accounts for both short-term dynamics (through \(p, d, q\)) and seasonal patterns (through \(P, D, Q\)) \citep{hyndman_forecasting_2018}. For example, a SARIMA(1,1,1)$\times$(1,1,1)\(_{12}\) model applies one level of regular and seasonal differencing, includes one AR and MA term at both the seasonal and non-seasonal levels, and assumes a yearly seasonality in monthly data.

We employ Seasonal ARIMA (SARIMA) models to forecast infection rates for different U.S. regions within 12 weeks of the start of vaccination. Given the biweekly temporal resolution of the data, we consider seasonal periods ($m$) ranging from 1 to 8 (corresponding to 2 to 16 weeks), allowing the model to flexibly detect short- or long-term seasonal trends. We perform an exhaustive grid search over model orders $(p,d,q)$ and seasonal orders $(P,D,Q)$ in $\{0,1\} \times \{1\} \times \{0,1\}$ and $\{0,1\}^3$, respectively. For each candidate seasonal period $m$, we fit the models and select the configuration that minimizes the Akaike Information Criterion (AIC). This model selection strategy balances fit and complexity and is well-established for univariate time series forecasting, where cross-validation is challenging. The resulting SARIMA models capture both the short-term autocorrelation and periodic epidemic waves across states.

In Table \ref{fig:validation-sarima}, we present the results of SARIMA for different values within the confidence interval predicted by automatic parameter selection. The seasonality factor $m$ represents the number of periods in a single seasonal cycle (here, two weeks). We determine the optimal seasonal period
$m$ for the SARIMA modeling by evaluating candidate values and selecting the one that minimizes the Akaike Information Criterion (AIC). This is done by searching the grid on SARIMA hyperparameters for each $m$, ensuring that the most fitting and parsimonious model is identified.

The $p-value$ scores are the results of a two-tailed t-test to compare the mean predicted value with the actual values, including the expected underreported cases. The range and average of infections are for the maximum and minimum infection rates within the confidence interval provided by SARIMA, for which the problem is feasible given the vaccination and current infected cases. We can also see the percentage of difference in the total number of infections.

Our results show that the quality of the SARIMA predictions is highly dependent on the quality of the data, and it is very sensitive to noise. For southern states, particularly, SARIMA did a poor job of predicting the infection rate. Furthermore, the infection rate is sensitive to change in policy, and solely temporal dependencies might not be enough to represent the patterns in the data. In this situation, shorter prediction periods and more frequent data acquisition seem to improve the results.


{\scriptsize
\begin{longtblr}[
    caption={Results of SARIMA analysis in different U.S. regions}, 
    label={fig:validation-sarima}]{
  colspec={>{\centering\arraybackslash}p{2cm}
           >{\centering\arraybackslash}p{2cm}
            >{\centering\arraybackslash}p{2.5cm}
           >{\centering\arraybackslash}p{2.5cm}
           >{\centering\arraybackslash}p{1.8cm}
           >{\centering\arraybackslash}p{2.1cm}
}}
\hline[1pt]
Region & Past Data Range& States with \textit{p-value} less than 0.025  & Periods with \textit{p-value} less than 0.025 &Infections (cases) & Difference with Actual Values (\%) \\
\hline
Midwest & 0--24 & IL, WI &0, 1, 2& 3,652,977 & 6.7\\
Midwest & 4--24 & IN, KS, ND &0, 1, 2& 1,634,226 & 58.3\\
Southwest & 0--24 & -&-& Infeasible & -\\
New England \& Middle Atlantic  & 0--24 & MA, ME, NH &0, 1, 2&3,145,257 & 8.2\\
West & 0--24 &  HI, MT, WY &-& 4,485,994 & 4.9 \\
\hline[1pt]
\end{longtblr}
}

\begin{figure}[h]
\vspace{-0.5cm}
\begin{minipage}{0.45\linewidth}
    \centering
        \includegraphics[width=\linewidth]{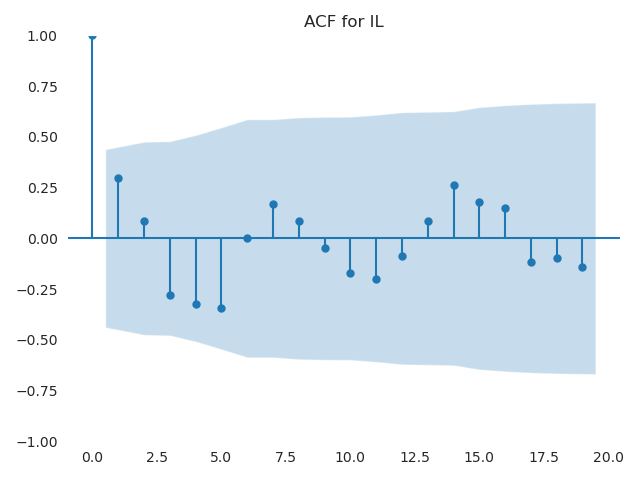}
        \caption{\centering ACF Diagram for Illinois}
    \label{appendix-fig:acf}
\end{minipage}
\hfill
\begin{minipage}{0.55\linewidth}
        \includegraphics[width=\linewidth]{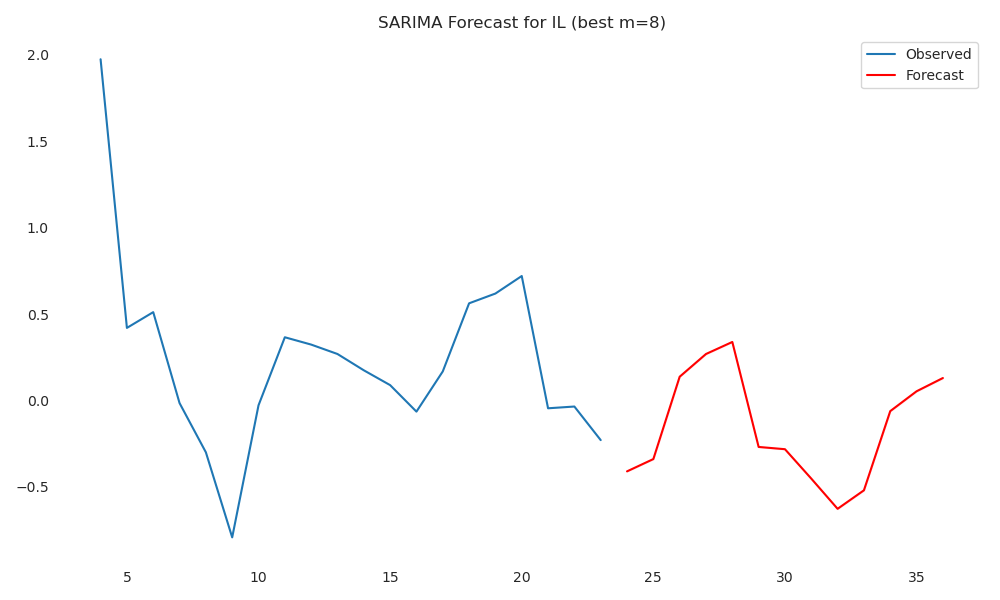}
        \caption{\centering SARIMA Time-series Prediction for Illinois}
        \label{appendix-fig:sarima}
\end{minipage}
\end{figure}

Figure \ref{appendix-fig:acf} displays the autocorrelation function (ACF) only for Illinois, but similar behavior in other states reveals the strength and significance of temporal correlations at various lags. Across most states, a sharp decline in autocorrelations is evident after the initial lags, with significant spikes at lag one and a gradual decrease thereafter, suggesting short-term dependencies. Some states show persistent negative autocorrelations at intermediate lags, indicative of potential oscillatory behavior or periodicity, whereas some exhibit weaker overall correlations beyond the first few lags, signaling relatively random residual behavior. In general, temporal dependencies vary across states, with most trends being short-lived. Using the selected seasonality parameters by AIC, we implement SARIMA to predict the unknown infection rate using the data from March 2020 to November 2020 for the following six months for different regions; Figure \ref{appendix-fig:sarima} depicts the results of this analysis for the state of Illinois. 

Our numerical experiments reveal that adjustments in SARIMA parameters significantly affect prediction accuracy, highlighting the importance of correctly identifying seasonal periodicity in SARIMA modeling. Although some differences were observed in the results of SARIMA compared to actual values, it can be a promising starting point for assessing the predictive power of our model. The projection using SARIMA will be more exact with the acquisition of more data; however, even in the initial phases of the pandemic, when the data is not abundant, it can be used to provide a confidence interval for the infection rate. Given the importance of correct predictions, it is worthwhile to discuss this case. As an example, we can consider when we allocate vaccines according to the projected infection rates; we might jeopardize human lives by using inaccurate estimates. One solution to the data scarcity problem can be projecting to short-term future periods and iteratively correcting the estimates. Another approach can be using regression models to predict the contributing factors to the seasonality of infections using similar precedent epidemics; the resulting models can be calibrated and used to calculate the seasonality parameter in the current disease based on known information. A robust stochastic analysis can be another good alternative when the resources are not too limited and we have different estimates of the seasonality parameter. In this case, we can use the worst-case scenario of infections as a baseline model to allocate resources in a risk-averse fashion.

{These patterns emphasize that model accuracy is both spatially and temporally heterogeneous, with higher reliability in the short term and in more stable regions.} This underscores the importance of high-quality and timely data for improving long-term forecasts. {Our findings suggest that more advanced prediction techniques, such as neural networks, may better capture nonlinear dynamics and complex dependencies in future work. However, while deep learning methods offer greater modeling power, they require extensive data and computational resources to match or surpass the performance of well-calibrated statistical models} (\cite{chollet2021deep}).\looseness-1

\section{Discussion on Results}\label{appendix: Appendix 6}

\noindent This section extends the discussion on the results of our model, as briefly provided in the main paper. We first provide the results of numerical experiments on Gini-based Formulation \eqref{objective1-eq1.1}--\eqref{eq6.10}. Then, we proceed to a comparative analysis of the model results with reality to investigate the possibility of improvements in vaccination campaigns. \looseness-1

\subsection{Gini-based Formulation}\label{appendix: Appendix 6-1} 

\noindent In this section, we present the results of our numerical experiments for Gini-based Formulation \eqref{objective1-eq1.1}--\eqref{eq6.10} and the modified spatiotemporal heuristic decomposition (Gini-based Decomposition \eqref{appendix-dec1-objective1-eq1.1}--\eqref{appendix-dec1-eq3}).
In this table, we consider the normalized version of Gini-based Formulation \eqref{objective1-eq1.1}--\eqref{eq6.10} and Gini-based Decomposition (\nameref{appendix: Appendix 3}, eqs. \eqref{appendix-dec1-objective1-eq1.1}--\eqref{appendix-dec1-eq3}). 
Table~\ref{tab:form1NHMA} reports results for the normalized Gini-based Formulation~\eqref{objective1-eq1.1}--\eqref{eq6.10} and Gini-based Decomposition (\nameref{appendix: Appendix 3}, eqs. \eqref{appendix-dec1-objective1-eq1.1}--\eqref{appendix-dec1-eq3}). 
Here, the optimality gap is calculated with reference to the Incumbent Obj., which is the best-known solution to the original MIP \eqref{objective1-eq1.1}--\eqref{eq6.10}, used in lieu of the true optimum due to intractability. \looseness-1


{\scriptsize
\begin{longtblr}[
    caption={Results for Gini-based Formulation~\eqref{objective1-eq1.1}--\eqref{eq6.10} in New England and Middle Atlantic Regions}, 
    label={tab:form1NHMA}]{
  colspec={>{\centering\arraybackslash}p{3.7cm}
           >{\centering\arraybackslash}p{1.5cm}
           >{\centering\arraybackslash}p{1.4cm}
           >{\centering\arraybackslash}p{1.4
           cm}
           >{\centering\arraybackslash}p{1.4cm}
           >{\centering\arraybackslash}p{1.4cm}
           >{\centering\arraybackslash}p{2cm}}
}
\hline[1pt]
Solution Methodology & Infections (cases) & MIP Gap(\%) & Opt. Gap(\%) & Sol. Time (s) & Gini Index \eqref{eq4.5} & Infections Averted \\
\hline
Normalized Gini-based Formulation~\eqref{objective1-eq1.1}--\eqref{eq6.10} & 4,196,602 & 0.04 & -- & 3,600 & 0 & 392,254 \\
Gini-based Decomposition \eqref{appendix-dec1-objective1-eq1.1}--\eqref{appendix-dec1-eq3}, weights $(-1,1)$ &4,169,585 & 0 & 0.6 &3.3  & 0 & 419,270 \\
\hline[1pt]
\end{longtblr}
}

As shown in Table~\ref{tab:form1NHMA}, the Gini-based Formulation~\eqref{objective1-eq1.1}--\eqref{eq6.10} was solved in 3,600 seconds with an 0.04\% MIP gap. In contrast, the Gini-based Decomposition~\eqref{appendix-dec1-objective1-eq1.1}--\eqref{appendix-dec1-eq3} achieved a 0.6\% optimality gap in under 4 seconds. Both approaches yield a zero Gini coefficient, indicating full equity across states, as enforced by the decomposition subproblem objective~\eqref{appendix-dec1-objective2-eq1.1}. The full formulation averted 392,254 infections, while the decomposition averted 419,270 infections--denoting a difference of 27,0165 cases. Infections averted are computed as the difference between observed cases ($4,588,856$) and those optimized by each model or solution method.

\subsection{Sensitivity Analysis on Key Parameters}\label{appendix: Appendix 6-3} 
In this section, we provide a visualization of the sensitivity analysis results.
\begin{figure}[h]
    \centering
    \includegraphics[width=1\linewidth]{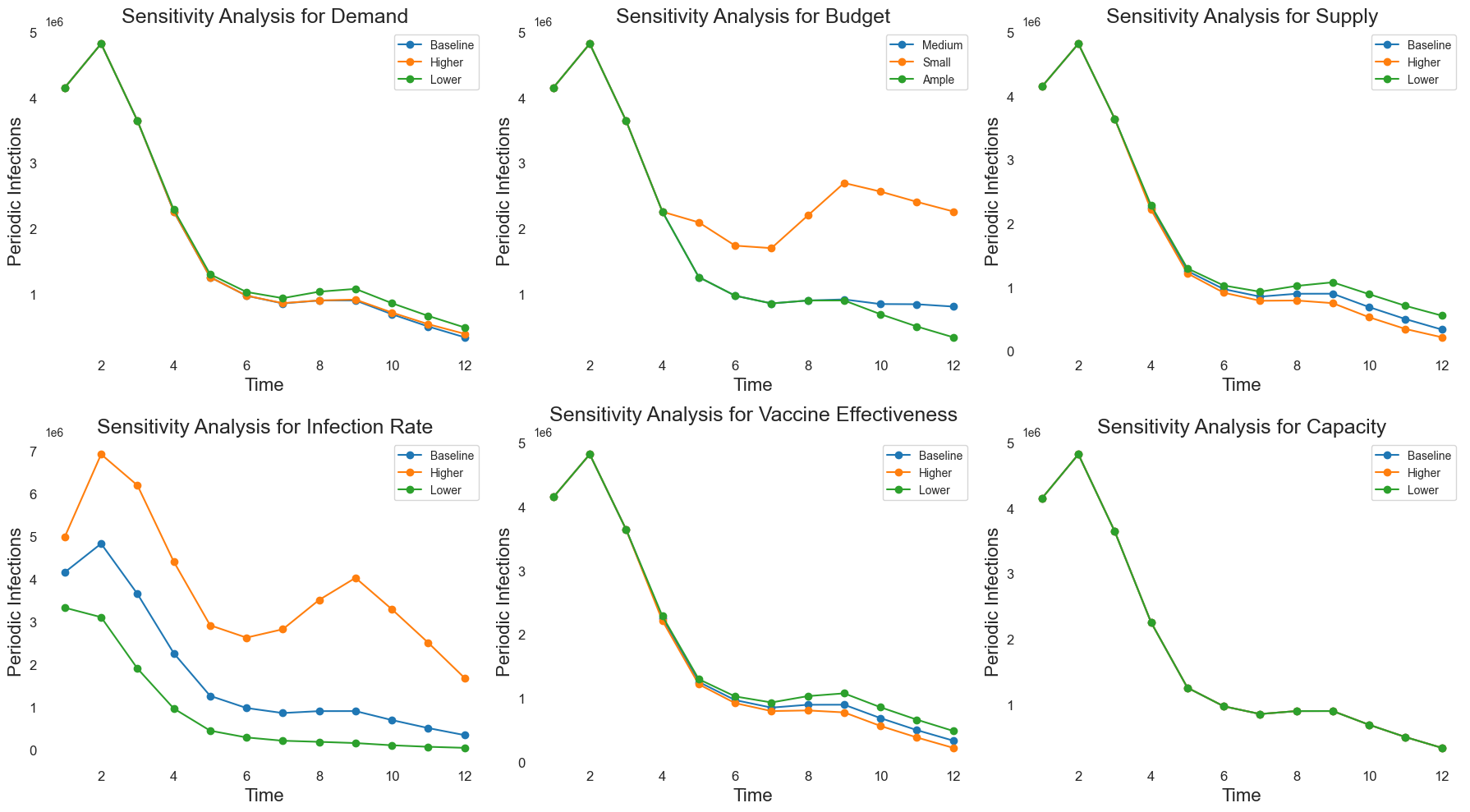}
    \caption{\centering Visualization of the Results of Sensitivity Analysis}
    \label{appendix-fig:sensitivity analysis}
\end{figure}
 Each subplot in Figure~\ref{appendix-fig:sensitivity analysis} shows the impact of changing one parameter—budget, supply, infection rate, vaccine effectiveness, capacity, and demand—on the infections over time. The baseline scenario corresponds to actual observed parameters. {Lines labeled 'Higher' and 'Lower' represent sensitivity variations aligned with the parameter changes detailed in the main text.} {Infection rate (top right) demonstrates the most dramatic impact, altering system behavior early and persistently. Other parameters influence outcomes more gradually, with noticeable effects emerging in later periods. This aligns with the dynamic, stock-dependent structure of epidemic models.}\looseness-1

\renewcommand\refname{References} 
{\small\bibliography{_Final_Submission}}

\end{document}